\documentclass[12pt]{article}
\usepackage{amssymb}

\makeatother

\def\theequation{\thesection.\arabic{equation}}

\def\q{\,\qquad}

\def\K{{\cal K}}
\def\W{{\cal W}}
\def\T{{\cal T}}

\def\pa{{\parallel}}
\def\pe{{\perp}}
\def\bp{{\mathbf{p}}}
\def\bt{{\mathbf{t}}}

\def\be{\begin{equation}}
\def\ee{\end{equation}}
\def\bea{\begin{eqnarray}}
\def\eea{\end{eqnarray}}

\def\ci  {{\mathcal I}}

\newcommand{\D}{{\mathcal{D}}}
\newcommand{\ie}{{\it i.e.,\,\,}}
\newcommand{\dr}{{\rm d}}
\newcommand{\cs}{\mathcal S}
\newcommand{\rhs}{{\it r.h.s.} }
\newcommand{\lhs}{{\it l.h.s.} }
\newcommand{\lhss}{{\it l.h.s.'s} }
\newcommand{\ta}{{\tau} }
\newcommand{\G}{{\mathcal G} }

\begin{document}

\begin{flushright}
\vspace{1mm}
 FIAN/TD/01--25\\
\vspace{-1mm}
\end{flushright}\vspace{1cm}

\makeatletter \@addtoreset{equation}{section} \makeatother
\def\theequation{\thesection.\arabic{equation}}

\begin{center}
{\large\bf Supersymmetric Higher-Spin Gauge Theories
in any $d$  and their\\ Coupling Constants within BRST Formalism
}
\vglue 0.6  true cm
\vskip0.8cm
{M.~A.~Vasiliev}
\vglue 0.3  true cm

I.E.Tamm Department of Theoretical Physics, Lebedev Physical Institute,\\
Leninsky prospect 53, 119991, Moscow, Russia
\vskip1.3cm
\end{center}

\vspace{1.2cm}

\vspace {1cm}

{\it $\phantom{MMMMMMMMMMMMMMMMMM}$ To the memory of Stanley Deser}


\vspace{0.4 cm}

\vspace{0.4 cm}

\begin{abstract}
Nonlinear field equations for the supersymmetric higher-spin gauge theory describing totally symmetric bosonic and fermionic massless fields  along with hook-type bosonic fields of all spins  in any space-time dimension are
presented. One of the novel features of the proposed formalism is that the
$osp(1,2)$ invariance and factorisation conditions are  formulated
within the BRST formalism,  that greatly simplifies the form of nonlinear
HS equations.
To match the list of vertices found by Metsaev, higher-spin gauge theory
is anticipated to possess an infinite number of independent coupling constants.
A conjecture that these coupling constants result from the locality restrictions on the elements of the factorisation ideal is put forward.

\end{abstract}

\newcommand{\su}{{{ hu(1,1|sp(2)[M,2]) }}}
\newcommand{\hc}{{{ hc(1|2\!\!:\!\![M,2]) }}}
\newcommand{\hco}{{{ hc(1|(1,2)\!\!:\!\![M,2]) }}}
\newcommand{\hu}{{{ hu(1|2\!\!:\!\![M,2]) }}}
\newcommand{\hupm}{{{ hu^E_\pm(1|(1,2)\!\!:\!\![M,2]) }}}
\newcommand{\huo}{{{ hu(1|(1,2)\!\!:\!\![M,2]) }}}
\newcommand{\hue}{{{ hu^E(1|(1,2)\!\!:\!\![M,2]) }}}
\newcommand{\hunmpq}{{{ hu(n,m|(u,v,2)\!\!:\!\![M,2]) }}}
\newcommand{\hunm}{{{ hu(n,m|(0,1,2)\!\!:\!\![M,2]) }}}
\newcommand{\honm}{{{ ho(n,m|(0,1,2)\!\!:\!\![M,2]) }}}
\newcommand{\huspnm}{{{ husp(n,m|(0,1,2)\!\!:\!\![M,2]) }}}
\newcommand{\honmpm}{{{ ho_\pm(n,m|sp(2)|[M,2]) }}}
\newcommand{\huspnmpm}{{{ husp_\pm (n,m,|sp(2)|[M,2]) }}}
\newcommand{\hunmpm}{{{ hu_\pm(n,m|(0,1,2)\!\!:\!\![M,2]) }}}
\newcommand{\supm}{{{ hu_\pm(1,1|sp(2)[M,2]) }}}
\newcommand{\ty}{\hat{y}}
\newcommand{\bee}{\begin{eqnarray}}
\newcommand{\eee}{\end{eqnarray}}
\newcommand{\nn}{\nonumber}
\newcommand{\lis}{Fort1,FV1,LV}
\newcommand{\hy}{\hat{y}}
\newcommand{\by}{\bar{y}}
\newcommand{\bz}{\bar{z}}
\newcommand{\go}{\omega}
\newcommand{\e}{\epsilon}
\newcommand{\f}{\frac}
\newcommand{\p}{\partial}
\newcommand{\half}{\frac{1}{2}}
\newcommand{\ga}{\alpha}
\newcommand{\gal}{\alpha}
\newcommand{\U}{\Upsilon}
\newcommand{\C}{{\bf C}}
\newcommand{\Y}{{\mathcal Y}}
\newcommand{\ups}{\upsilon}
\newcommand{\bu}{\bar{\upsilon}}
\newcommand{\dga}{{\dot{\alpha}}}
\newcommand{\dgb}{{\dot{\beta}}}
\newcommand{\gb}{\beta}
\newcommand{\gga}{\gamma}
\newcommand{\gd}{\delta}
\newcommand{\gl}{\lambda}
\newcommand{\gk}{\kappa}
\newcommand{\gep}{\epsilon}
\newcommand{\gvep}{\varepsilon}
\newcommand{\gs}{\sigma}
\newcommand{\V}{|0\rangle}
\newcommand{\ws}{\wedge\star\,}
\newcommand{\gee}{\epsilon}
\newcommand\ul{\underline}
\newcommand\un{{\underline{n}}}
\newcommand\ull{{\underline{l}}}
\newcommand\um{{\underline{m}}}
\newcommand\ur{{\underline{r}}}
\newcommand\us{{\underline{s}}}
\newcommand\up{{\underline{p}}}
\newcommand\A{{\mathcal A}}
\newcommand\B{{\mathcal B}}

\newcommand\uq{{\underline{q}}}
\newcommand\ri{{\cal R}}
\newcommand\punc{\multiput(134,25)(15,0){5}{\line(1,0){3}}}
\newcommand\runc{\multiput(149,40)(15,0){4}{\line(1,0){3}}}
\newcommand\tunc{\multiput(164,55)(15,0){3}{\line(1,0){3}}}
\newcommand\yunc{\multiput(179,70)(15,0){2}{\line(1,0){3}}}
\newcommand\uunc{\multiput(194,85)(15,0){1}{\line(1,0){3}}}
\newcommand\aunc{\multiput(-75,15)(0,15){1}{\line(0,1){3}}}
\newcommand\sunc{\multiput(-60,15)(0,15){2}{\line(0,1){3}}}
\newcommand\dunc{\multiput(-45,15)(0,15){3}{\line(0,1){3}}}
\newcommand\func{\multiput(-30,15)(0,15){4}{\line(0,1){3}}}
\newcommand\gunc{\multiput(-15,15)(0,15){5}{\line(0,1){3}}}
\newcommand\hunc{\multiput(0,15)(0,15){6}{\line(0,1){3}}}
\newcommand\ls{\!\!\!\!\!\!\!}

\newpage
\tableofcontents
\newpage

\section{Introduction}

Higher-spin (HS) gauge theories are fascinating theories of gauge fields of all
spins (see e.g. \cite{Vasiliev:1999ba,Bekaert:2004qos,Ponomarev:2022vjb} for  reviews), that may correspond to
most symmetric vacua of a theory of fundamental interactions
presently identified with superstring theory. Stanley Deser made a fundamental contribution
into the variety of directions in HS theory. In particular, in collaboration with Aragone,
he has shown that HS gauge theories admit no consistent gravitational
interaction in the flat background \cite{Aragone:1979hx}, and proposed a vierbein (frame-like) formulation
for HS fermion fields  \cite{Aragone:1980rk} which simultaneously was
proposed  both for bosons and for fermions of all spins in
\cite{Vasiliev:1980as}. The list of remarkable achievements of Stanley in HS theory
extends to $3d$ HS gauge theories
\cite{Deser:1981wh}-\cite{Aragone:1987dtt}, massive and partially massless fields \cite{Deser:1983tm}-\cite{Deser:2001us} and many other results in gravity \cite{Arnowitt:1959ah}, supergravity \cite{Deser:1976eh} and beyond.
Stanley Deser was great scientist, one of the leaders in the field of HS theory for many years also prominent for numerous other scientific achievements. On the top of that he was a man of mark with dramatic fate \cite{book}.

The characteristic feature of HS gauge theory is that it must respect rich HS gauge
symmetries.
Hence, the problem is to introduce interactions of  HS fields
in a way compatible with  nonabelian
HS gauge symmetries containing diffeomorphisms and Yang-Mills
symmetries as their parts.
Full nonlinear dynamics of HS gauge fields
has been  elaborated  at the level of equations of motion
for $d=4$ \cite{more}, which is the simplest nontrivial case
since HS gauge fields do not propagate if $d<4$, and for any $d$
in \cite{Vasiliev:2003ev}.
From the lower-order analysis of interactions of HS gauge fields in the framework
of gravity worked out at the action level for $d=4$  \cite{FV1} it was
found that\\
(i) consistent HS theories contain infinite sets of
infinitely increasing spins;\\
(ii) HS gauge
interactions contain higher derivatives;\\
(iii) in the framework of gravity, unbroken
HS gauge symmetries require a non-zero cosmological constant;\\
(iv)  HS symmetry algebras
\cite{FVa} are certain star-product algebras
\cite{Fort2}.

The properties (i) and (ii) were deduced in the remarkable
earlier works  \cite{Bengtsson:1983pd,Berends:1984wp} on  HS interactions in flat space.
The feature that unbroken
HS gauge symmetries require a non-zero cosmological constant \cite{FV1}
is  crucial in several respects,
explaining in particular why the analysis of HS--gravitational interactions
in the framework of the expansion near the flat
background led to the negative conclusions in \cite{Aragone:1979hx}.
The same time it fits the idea of  holographic correspondence between
HS gauge theories in the bulk and boundary conformal theories
\cite{Sun}-\cite{KP}.

HS theories were explored within various approaches (for the incomplete list of references see, {\it e.g.},
\cite{Bengtsson:1983pd,Berends:1984wp,Berends:1984rq,FV1}, \cite{Metsaev:1991mt}-\cite{Buchbinder:2025ceg},
where cubic HS vertices were studied by a number of formalisms in the lowest order, that does not determine the coupling constants). Very important results were obtained by Metsaev who in particular obtained
the full classification of the cubic $P$-even HS vertices in Minkowski space of any dimension $d\geq 4$ in
\cite{Metsaev:2005ar,Metsaev:2007rn}.

In the recent paper \cite{Tatarenko:2024csa} it has been checked that at $d=4$ the vertices classified by Metsaev
precisely match the current deformation of the free HS equations
resulting from the non-linear $4d$
HS theory of \cite{more}. (For the related preceding work see also \cite{Misuna:2017bjb}.)
 Namely, according to \cite{Metsaev:2005ar}, at $d=4$
there are vertices associated with the two types of currents for any three spins $s_{1,2,3}$
 $$
 J^{min}_{s_1,s_2,s_3} :\qquad N^{max}_{der}\leq  s_1+s_2+s_3 - 2 s_{min}\,,
 $$
$$
\ls\ls \ls\ls J^{max}_{s_1,s_2,s_3} :\qquad N^{max}_{der}\leq  s_1+s_2+s_3 \,,
 $$
 where $N^{max}_{der}$ is the minimal possible number of
 maximal derivatives in the current. (In $AdS$, currents  also contain subleading
 derivative terms with the coefficients proportional to the powers of the cosmological constant.)
  Since fields of all spins form a multiplet of the HS algebra, the spin-dependent coefficients in front of different currents are determined in terms of the two independent coupling constants of the non-linear $4d$ HS theory of \cite{more}.
  Let us stress that, relating fields and currents of different spins, HS symmetries do not relate the currents of different types. This is why the two independent
  couplings survive in the nonlinear HS theory.

For $d>4$ the list of cubic vertices found by Metsaev
\cite{Metsaev:2007rn} is different. Namely, for any
spins $s_1$, $s_2$, $s_3$,
there are vertices   with various maximal numbers of derivatives in the interval
\be
 N^{max}_{der} = s_1+s_2+s_3 - 2n\,, \q 0\leq n\leq s_{min}\,.
\ee
Since the number of independent couplings (currents) increases with spins while the full nonlinear
HS theory contains infinite towers of spins the latter is anticipated to possess an infinite
number of independent coupling constants. On the other hand, the HS model of \cite{Vasiliev:2003ev} has only
one coupling constant. This raises the questions whether the HS gauge theory in arbitrary dimension
admits a generalization rich enough to incorporate all couplings of  Metsaev's classification. One of the  goals of this paper is to conjecture  a
mechanism for such a generalization.

The idea is the following. The construction of HS theory of \cite{Vasiliev:2003ev}
contains the factorisation of elements of the form $\ta_{ij} * f^{ij}=f^{ij}* \ta_{ij}$ where $\ta_{ij}$ are certain $sp(2)$
generators (for detail see  Section \ref{refine}). This factorisation puts the system on-shell effectively taking away
all terms proportional to the D'Alembertian. Such a procedure is however ambiguous unless the functional class $F$
of elements $f^{ij}$ is specified. In the old days of  \cite{Vasiliev:2003ev} not so much  information (if any) was available allowing to choose
 the appropriate class. The situation has changed during several last years
as a result of the analysis of the issue of locality in HS theory. The two new key notions are  spin-locality
\cite{Gelfond:2018vmi} and projective compactness of vertices \cite{Vasiliev:2022med}. (See also Appendix A.)
In particular, it has been shown in \cite{Vasiliev:2022med} that the field redefinitions in the HS theory that
preserve spin-locality and make equivalent these  concepts both in space-time
 and in the auxiliary fiber space belong to the
projectively-compact spin-local class being associated with $F$
in this paper. In other words, if a $f^{ij} \notin F$ it should
not contribute to the factorisation process. As a result, many vertices treated as trivial in \cite{Vasiliev:2003ev}, may in fact survive because their compensation (field redefinition) procedure was not spin-local projectively-compact.

In this paper we show that the modified setup does not affect the free field analysis, that is an important consistency check.  The more involved details
of the nonlinear analysis are  postponed for a future publication.

Another goal is to introduce a new class of supersymmetric HS (SHS)
theories in any dimension  involving both bosons and fermions. (Note that, being supersymmetric in the
HS sense, in higher dimensions these models may not be supersymmetric in the standard sense
since space-time (super)generators do not form its proper subalgebra.) A class of couplings, that
hopefully  resolve the seeming conflict with Metsaev's vertex classification
in these models is proposed as well.
Note that, as discussed in Section \ref{nonlsup}, the construction of the SHS models has a number of tricky points in the fermionic sector having no counterparts in the bosonic case.

A more technical but  important new element of the proposed formalism is the realization of the $sp(2)$ ($osp(1,2)$ in the supersymmetric case) within the
BRST technique. By virtue of additional variables associated with the BRST ghosts, this approach makes the full nonlinear system of equations as simple as the $4d$ system of \cite{more}. Interestingly enough it automatically puts it on shell.

As a byproduct we observe that the proposed approach has much in common with the
BRST approach to String Theory providing a promising tool for the unification of
HS theory and String Theory via association of the BRST operator $Q$ with $2d$ CFTs.

The layout of the rest of the paper is as follows.

In Section \ref{A} we recall the $A$-model HS equations of
\cite{Vasiliev:2003ev}.

Some elementary facts of the BRST approach are recalled in Section
 \ref{brst} with the emphasis on the distinction between the left
 and adjoint actions of the BRST operator.

In Section \ref{refine} the $A$-model is reformulated in a novel
form allowing to specify a class of
functions in which the factorisation parameters are valued. In particular,  in Section \ref{lina} it is shown that the new setup for
 formulating the $sp(2)$ invariance and factorisation conditions in terms of the BRST operator leads to usual linearized HS equations.
In this section a conjecture is put forward  that the vast variety of the coupling constants in the theory should result from the restriction of the parameters of the factorisation  transformations to the projectively-compact spin-local class.

 In Section \ref{SHSasg}, the supersymmetric HS algebras of \cite{Vasiliev:2004cm} are reformulated in terms of Clifford variables most convenient for the formulation of the nonlinear theory. Some useful relations in $U(osp(1,2))$ are presented in Section \ref{osp}.

 The nonlinear SHS field equations are formulated in Section \ref{nonlsup}. The extension to the SHS  model is not trivial in several  respects and, first of all, in the proof of $osp(1,2)$ invariance on the dynamical fields
where  the BRST formalism again plays the key role. The linearised analysis is shown to reproduce anticipated free unfolded equations in Section \ref{pertshs}
 while the inner symmetry
 extensions  are considered in Section \ref{inn}.

 In Section \ref{disc} some conclusions and perspective are discussed with the
 emphasis on the new elements of the construction of this paper and potential
 implications on the holographic interpretation of the conjecture on the variety of the coupling constants of the HS theory. Possible links between (S)HS gauge theory and String Theory are briefly considered.

 The key ingredients of the concepts of spin-locality and projective compactness are sketched in Appendix A. Appendix B presents detail of the equivalence proof of
 the $osp(1,2)$ invariance and factorisation conditions within the BRST free formulation.

\section{
Original form of  type\,-A   higher-spin gauge theory}
\label{A}

In this section  the construction of the so-called type--$A$ HS gauge theory of \cite{Vasiliev:2003ev} is recalled.

\subsection{Free fields}
\label{Free Fields}

In the frame-like formalism initiated in
\cite{Aragone:1980rk,Vasiliev:1980as}, a spin $s$ gauge field in $AdS_d$ is conveniently described
 by a one-form $ \go{}^{A_1 \ldots A_{s-1}, B_1\ldots B_{s-1} }$
valued in the irreducible representation of  $o(d-1,2)$
$ (A,B = 0,\ldots , d)$ described by the traceless two-row
rectangular Young diagram of length $s-1$
\be
\label{irre}
\go^{\{A_1 \ldots A_{s-1},A_s\} B_2\ldots B_{s-1} } =0\,,\qquad
\go^{A_1 \ldots A_{s-3}C}{}_{C,}{}^{B_1\ldots B_{s-1} } =0\,.
\ee
(For more detail we refer the reader to the original papers \cite{LV,5d}
and review \cite{Bekaert:2004qos}.)

For instance, the spin-two field of $d$-dimensional gravity is described
 by a one-form connection  $\go^{AB}= - \go^{BA}$
of the $(A)dS_d$ Lie algebra  $o(d-1,2)$.
The Lorentz subalgebra $o(d-1,1)$  is a
stability subalgebra of some vector $V^A$, that can be chosen
differently in different points of space-time, thus becoming a
field  $V^A = V^A (x)$. Its norm  is convenient
to relate to the cosmological constant
$\Lambda$ so that $V^A$ has dimension of length
\be
\label{vnorm}
V^AV_A = -\Lambda^{-1}\,.
\ee
$\Lambda$ is  negative  in $AdS_d$ with  mostly minus
signature. This makes it possible to give a covariant definition of the
frame field and Lorentz connection \cite{SW,PrV}
\be
\label{defh}
 E^A = D(V^A) \equiv \dr_x V^A + \go^{AB}V_B\,,\qquad
\go^{L\,AB} = \go^{AB} +\Lambda  ( E^A V^B - E^B V^A )\,.
\ee
According to these definitions
$
E^A V_A =0\,,
$
$
D^L V^A = dV^A + \go^{L\,AB}V_B \equiv 0\,.
$
When the  frame $E_\un^A$ has the maximal rank $d$ it
gives rise to a nondegenerate metric tensor
$
g_{\un\um} = E_\un^A E_\um^B \eta_{AB}\,
$
in the $d$-dimensional space. The torsion two-form is
$
r^A:= DE^A\equiv  r^{AB} V_B \,.
$
The zero-torsion condition
$
r^A = 0\,
$
expresses the Lorentz connection via
derivatives of the frame field in a usual manner. The
$V^A$ transversal components of the curvature (\ref{Rvac}) $r^{AB}$
identify with the Riemann tensor shifted by the term  bilinear in the
frame one-form. As a result, any field $\go_0$ satisfying
the zero-curvature equation
\be
\label{Rvac}
r^{AB}= \dr_x \go_0^{AB} + \go_0^{A}{}_C \go_0^{CB}=0\,
\ee
describes locally $(A)dS_d$ space-time
with the cosmological term $\Lambda$ provided that the metric
tensor is nondegenerate. (Note that in this paper we ignore the wedge symbol $\wedge$ since all products of differential forms are exterior.)

The Lorentz irreducible HS connections
$ \go{}^{a_1 \ldots a_{s-1}, b_1\ldots b_t }$ originally introduced in
\cite{Vasiliev:1980as,LV} are the $d$-dimensional traceless parts of
those components of $ \go{}^{A_1 \ldots A_{s-1}, B_1\ldots B_{s-1} }$
that are parallel to $V^A$ in $s-t-1$  indices and transversal in
the rest ones. Let some solution to
(\ref{Rvac}), that describes the $(A)dS_d$ background, be fixed.
The linearized HS curvature $R_1$ of the form
\bee
\label{R1A}
R_1^{A_1 \ldots A_{s-1}, B_1\ldots B_{s-1} } &=& D_0
(\go_1^{A_1 \ldots A_{s-1}, B_1\ldots B_{s-1}}) :=
d \go_1^{A_1 \ldots A_{s-1}, B_1\ldots B_{s-1} }\nn\\
 &{}&\ls\ls\ls\ls\ls\ls\ls\ls\ls +(s-1)\Big(
\go_0^{\{A_1}{}_{C}\wedge
\go_1^{C A_2 \ldots A_{s-1}\}, B_1\ldots B_{s-1} }
+\go_0^{\{B_1}{}_{C}\wedge
\go_1^{ A_1 \ldots A_{s-1}, C B_2\ldots B_{s-1}\} }\Big )\,
\eee
 is manifestly invariant under the linearized HS gauge transformations
\be
\label{litr}
\delta \go{}^{A_1 \ldots A_{s-1}, B_1\ldots B_{s-1} }(x) =
D_0 \gvep {}^{A_1 \ldots A_{s-1}, B_1\ldots B_{s-1} } (x)
\ee
because, according to (\ref{Rvac}), $D_0^2 \equiv r(\go_0 )=0$.

\subsection{Bosonic higher-spin algebra}
\label{Oscillator Realization}

{}From Section \ref{Free Fields} it is clear that,
to reproduce  the correct set of HS gauge fields, one has to find such
an algebra $g$ that  contains $h=o(d-1,2)$ or  $h=o(d,1)$
 as a subalgebra and decomposes under the adjoint action
of $h$  into a sum of
irreducible finite-dimensional $h$-modules
 described by various two-row rectangular
traceless Young tableaux.
Such  algebra called usually type-$A$ HS algebra was described by Eastwood in \cite{Eastwood:2002su} as the algebra of conformal HS symmetries of
the free  massless Klein-Gordon equation in $d-1$ dimensions.
Here we give following \cite{Vasiliev:2003ev} its alternative realisation
 more suitable for the analysis of the HS interactions.

Consider oscillators $Y_i^A$ with $i=1,2$ satisfying the
commutation relations
\be
\label{defr}
[Y_i^A , Y_j^B ]_* = \gvep_{ij}\eta^{AB}\,,\qquad \gvep_{ij}= -
\gvep_{ji}\,,\quad \gvep_{12}=1\,,
\ee
where  $\eta^{AB}$ is the invariant symmetric  form of
$o(n,m)$. For example, one can interpret these
oscillators as conjugated coordinates and momenta, $Y^A_1 = P^A$,
$Y^B_2 = Y^B$.
$\eta^{AB}$ and $\gvep_{ij}$ ($\gvep^{ik}\gvep_{il}= \delta^k_l$) raise and
lower indices in the usual manner
$
A^A = \eta^{AB} A_B
$,
$
a^i =\gvep^{ij}a_j
$,
$
a_i =a^j \gvep_{ji}\,.
$

We use the Weyl (Moyal) star product
\be
\label{wstar}
(f*g)(Y) :=\f{1}{\pi^{2(d+1)}}
\int dS dT f(Y+S) g(Y+T)\exp -2 S^A_i T_A^i\,.
\ee
$[f ,g ]_* := f*g - g*f$, $\{f,g\}_* := f*g + g*f$.
The associative algebra of polynomials
with the star-product law generated via
(\ref{defr}) is called Weyl algebra $A_{d+1}$. Its generic
element is
$f(Y) = \sum \phi^{i_1 \ldots i_n}_{A_1 \ldots A_n} Y_{i_1}^{A_1}\ldots
Y_{i_n}^{A_n}\,$
or, equivalently,
\be
\label{exp}
f(Y) = \sum_{m,n} f_{A_1 \ldots A_m\,,B_1 \ldots B_n} Y_{1}^{A_1}\ldots
Y_{1}^{A_m}Y_{2}^{B_1}\ldots
Y_{2}^{B_n}\,
\ee
with the coefficients $f_{A_1 \ldots A_m\,,B_1 \ldots B_n}$
symmetric in the indices $A_i$ and $B_j$.

 With respect to star commutators,
various bilinears built from the oscillators $Y_i^A$ form the
Lie algebra $sp(2(d+1))$. It contains
the subalgebra $o(d-1,2)\oplus sp(2)$ spanned by the
mutually commuting generators
\be
\label{t}
T^{AB} = -T^{BA} :=\half  Y^{iA} Y^B_i\,,\qquad
t_{ij} =t_{ji} := Y^A_i Y_{jA}  \,.
\ee
Consider the subalgebra $S\subset A_{d+1}$
spanned by the $sp(2)$ singlets $f(Y)$,
\be
\label{sp2}
[t_{ij} , f(Y) ]_* =0\,.
\ee
Eq.(\ref{sp2}) yields
$
\Big(Y^{Ai} \f{\p}{Y^A_j}  + Y^{Aj} \f{\p}{Y^A_i}
\Big) f(Y) =0\,,
$
  which implies that the coefficients
$f_{A_1 \ldots A_m\,,B_1 \ldots  B_n}$ in (\ref{exp}) are nonzero
only if $n=m$
and that symmetrization over any $m+1$ indices of
$f_{A_1 \ldots A_m\,,B_1 \ldots  B_m}$ yields zero, \ie they
 have the symmetry properties of a two-row rectangular
Young diagram. As a result, the gauge fields of $S$ are
\be
\label{gaug}
\go(Y|x) = \sum_{l=0}^\infty \go_{A_1 \ldots A_l\,,B_1 \ldots B_l}(x)
 Y_{1}^{A_1}\ldots Y_{1}^{A_l}Y_{2}^{B_1}\ldots Y_{2}^{B_l}
\ee
with the component gauge fields $\go_{A_1 \ldots A_l\,,B_1 \ldots
B_l}(x)$  valued in all two-row rectangular Young diagrams of $gl(d+1)$
(no metric and, hence, tracelessness conditions are imposed so far).

Algebra $S$ is not simple,
containing the two-sided ideal $I$ spanned by the elements
of the form
\be
\label{I}
g=t_{ij}*g^{ij}\,,
\ee
where $g^{ij}$ transforms as a symmetric tensor with respect to
$sp(2)$,
\be
[t_{ij}\,,g^{kl}]_* = \delta_i^k g_j{}^l +\delta_j^k g_i{}^l
+\delta_i^l g_j{}^k +\delta_j^l g_i{}^k\,.
\ee
(Note that $t_{ij}*g^{ij} =
g^{ij} *t_{ij}$.)
Indeed, from (\ref{sp2}) it follows that
$f*g,\, g*f \in I$\, $\forall f\in S$, $g\in I$.
  Due to the definition (\ref{t}) of $t_{ij}$,
the ideal $I$ contains all traces
of the two-row Young tableaux, while the algebra
$S/I$ has only traceless two-row tableaux in the expansion (\ref{gaug}).

For the complex algebra $S/I$ we will use
notation $hgl(1|sp(2)[d+1])$. For the generalizations and real forms
corresponding to unitary HS theories  see Section \ref{inn} and  \cite{Vasiliev:2003ev}.

Note that the described construction  of the HS algebra
is analogous to that of the $AdS_7$ HS algebra given by
Sezgin and Sundell in  \cite{7}
 in terms of spinor oscillators with the symmetric $7d$
charge conjugation matrix in place of the metric tensor in
(\ref{defr}). Also  note that the key role of
the algebra $sp(2)$ in the analysis of HS dynamics explained below
is in many respects analogous to that of $sp(2)$ in the two-time
approach of Bars \cite{Bars}.

\subsection{Twisted adjoint module and central on-mass-shell theorem}
\label{Twisted representation and Central On-Mass-Shell Theorem}

In HS gauge theories, the construction of the twisted adjoint module, where the
HS Weyl zero-forms are valued, is based on such  involutive automorphism $\tau$
of the  HS algebra that
\be
\tau (P^a ) = -P^a\,,\qquad
\tau (L^{ab} ) = L^{ab}\,.
\ee
Once the Lorentz algebra is singled out by the compensator $V^A$, the
automorphism $\tau$ describes the reflection with respect to $V^A$. In particular, for the HS algebra under
investigation
$
\tau (Y^A_i) = \tilde{Y}^A_i\,,
$
where
\be
\label{tila}
\tilde{A}^A := A^A -\f{2}{V^2}V^A V_B A^B\,, \qquad \forall A^A\,.
\ee

Following \cite{Vasiliev:2003ev} we use notations
\be
A^{A}_i = {}^\pa A^{ A}_i + {}^\pe A^{ A}_i\,,\qquad
{}^\pa A^{ A}_i :=  \f{1}{V^2} V^A V_B A^B_i\,,\quad
{}^\pe A^{ A}_i :=  A^A_i - \f{1}{V^2} V^A V_B A^B_i\,
\ee
so that ${}^\pa \tilde{A}^A_i=-{}^\pa A^A_i$ and
${}^\pe \tilde{A}^A_i={}^\pe A^A_i$.
For a general element $f(Y)$,
\be
\label{tif}
\tau (f(Y )) = \tilde{f} (Y) := f (\tilde{Y} )\,.
\ee

Let $C(Y|x)$ be  a zero-form in the HS algebra vector space, \ie
$[t_{ij} , C]_* =0$ with the ideal $I$ factored out.
The covariant derivative in the twisted adjoint module is
\be
\tilde{D} (C)= \dr_x C + \go *C - C*\tilde{\go}\,.
\ee
(Note that $\tilde t^{ij}= t^{ij}$.)

Central On-Mass-Shell theorem  formulated in \cite{5d}
in terms of Lorentz irreducible components of $C(Y|x)$ states  that the Fronsdal
equations for totally symmetric free
massless fields in $(A)dS_d$ \cite{Fr,Fronsdal:1978vb}
supplemented by an infinite set of constraints, that express an infinite set
 of the auxiliary fields in terms of the Fronsdal fields and their derivatives,
can be formulated in the form
\be
\label{CMS1}
R_1 ({}^\pa Y , {}^\pe Y ) = \half E_0^A   E_0^B
\f{\p^2}{\p Y^A_i \p Y^B_j} \gvep_{ij} C(0,{}^\pe Y)\,,
\ee
\be
\label{CMS2}
\tilde{D}_0 (C) =0\,,
\ee
where
\be
\label{COMT1}
R_1 (Y) = \dr_x\go(Y) + \go_0*\go +\go * \go_0\,,
\ee
\be
\label{COMT2}
\tilde{D}_0 (C)= \dr_x C + \go_0 *C - C*\tilde{\go}_0
\ee
and $\go_0 := \go_0^{AB} (x) T_{AB}$ with the vacuum $AdS_d$ connection
$\go_0^{AB} (x)$  satisfying (\ref{Rvac}).
The theorem states that equations
(\ref{CMS1}), (\ref{CMS2}) are equivalent to the Fronsdal equations supplemented by an infinite set of algebraic constraints on auxiliary  fields.

The components of the expansion of the
zero-forms $C(0,{}^\pe Y)$ on the \rhs
of (\ref{CMS1}) in powers of $Y^A_i$
are $V^A$-transversal. These are the
HS Weyl zero-forms $C^{a_1 \ldots a_{s}, b_1\ldots b_{s} }$
described by the length $s$ traceless two-row rectangular Lorentz
Young diagrams. They  parameterize
those components of the HS field strengths that may be
non-zero when the field equations and constraints on extra fields
are satisfied. Equation (\ref{CMS2}) describes the
consistency conditions for the HS equations plus dynamical
equations for spins 0 and 1. (Dynamics of a massless scalar
was described this way in \cite{SHV}.)
In addition, they
express an infinite tower of auxiliary fields  contained
in $C$ in terms of derivatives of the dynamical (\ie Fronsdal)
HS fields.

The key fact is that  equations (\ref{CMS1}), (\ref{CMS2})
are consistent, \ie application of the covariant
derivative to the \lhss of (\ref{CMS1}), (\ref{CMS2})
does not lead to new conditions as is not hard to see directly (see also \cite{Bekaert:2004qos,Vasiliev:2003ev}).

 The covariant derivatives in the adjoint and
twisted adjoint representations have the form
\be
\label{ad}
D_0 := D^L_0 -\Lambda  E^A_0 V^B \Big ( {}^\pe Y_{Ai}
\f{\p}{\p {}^\pa Y^{ B}_i} -
{}^\pa Y_{Bi}\f{\p}{\p {}^\pe Y^A_i} \Big )\,,
\ee
\be
\label{twad}
\tilde{D}_0 := D_0^L - 2\Lambda  E_0^A V^B \Big ( {}^\pe Y_{A}^i {}^\pa
Y_{Bi} -
\f{1}{4}\gvep^{ji}
\f{\p}{ \p {}^\pe Y^{ A j}\p {}^\pa Y^{Bi }} \Big )\,,
\ee
where
\be
D_0^L := \dr_x +  \go_0^{L\, AB}
{}^\pe Y_{Ai}\f{\p}{\p {}^\pe Y^B_i}  \,.
\ee

The adjoint covariant derivative (\ref{ad}) and twisted adjoint
one (\ref{twad}) commute with the operators $N^{ad}$ and  $N^{tw}$, respectively,
\be
N^{ad}:= Y^A_i\f{\p}{\p Y^A_i}\,,\qquad
N^{tw}:= {}^\pe Y^A_i\f{\p}{\p {}^\pe Y^A_i}-
{}^\pa Y^A_i\f{\p}{\p {}^\pa Y^A_i}\,.
\ee
This implies that the free field equations (\ref{CMS1}) and
(\ref{CMS2}) decompose into independent subsystems
for the sets of fields of different spins  $s$ obeying
\be
\label{spin}
N^{ad} \go = 2(s-1) \go\,, \qquad N^{tw} C = 2s C\,, \qquad s \geq 0\,.
\ee
(Note that $N^{tw}$ has no negative eigenvalues
 on the  two-row rectangular Young diagram tensors   because
having more than a half of vector indices
aligned along  $V^A$ would imply symmetrization
over more than a half of indices,
thus giving zero.)

 In terms of the
Lorentz irreducible components, the spin $s$ gauge connections are
valued in the representations
\begin{picture}(43,15)
\put( 9,12){\scriptsize   s-1}    
\put( 9,-5){\scriptsize   t}    
{\linethickness{.250mm}
\put(00,10){\line(1,0){30}}
\put(00,05){\line(1,0){30}}
\put(00,00){\line(1,0){15}}
\put(00,00){\line(0,1){10}}%
\put(05,00.0){\line(0,1){10}}
\put(10,00.0){\line(0,1){10}}
\put(15,00.0){\line(0,1){10}}
\put(20,05.0){\line(0,1){05}}
\put(25,05.0){\line(0,1){05}}
\put(30,05.0){\line(0,1){05}}
}
\end{picture}
of \cite{LV} with various $0\leq t\leq s-1$ while
the spin $s$ Weyl tensors are valued in the Lorentz representations
\begin{picture}(43,15)
\put( 9,12){\scriptsize   p}    
\put( 9,-5){\scriptsize   s}    
{\linethickness{.250mm}
\put(00,10){\line(1,0){30}}
\put(00,05){\line(1,0){30}}
\put(00,00){\line(1,0){15}}
\put(00,00){\line(0,1){10}}%
\put(05,00.0){\line(0,1){10}}
\put(10,00.0){\line(0,1){10}}
\put(15,00.0){\line(0,1){10}}
\put(20,05.0){\line(0,1){05}}
\put(25,05.0){\line(0,1){05}}
\put(30,05.0){\line(0,1){05}}
}
\end{picture} with various $p\geq s$. (Note that the missed cells
 compared to the rectangular diagram of the length
of the upper row correspond to the Lorentz
invariant direction along $V^A$.) We observe that  the
twisted adjoint action of the $(A)dS_d$ algebra
decomposes into an infinite set of infinite-dimensional submodules
associated with different spins, while its adjoint action on the HS algebra
decomposes into an infinite set of finite-dimensional submodules.

\subsection{Nonlinear equations}
\label{Nonlinear Equations}

The key principle that allowed us to
build in \cite{Vasiliev:2003ev} bosonic HS equations in any $d$ is that in that case one has to demand
the existence of  the $sp(2)$ algebra at the nonlinear level, which singles out the
HS algebra spanned by two-row rectangular tensor elements. Otherwise,
the condition (\ref{sp2}) would not allow a
meaningful extension beyond the free field level, \ie the resulting
system may admit no interpretation in terms of the original HS
tensor fields described by the two-row rectangular Young diagrams, that would
lead to nonlinear equations on the
 tensors absent at the free level.


To this end, following \cite{Vasiliev:2003ev},  we double a number of oscillators
$Y^A_i \to (Z^A_i , Y^A_i )$, endowing the space of functions $f(Z,Y)$
with the associative star product
\be
\label{star}
(f*g)(Z,Y) :=\f{1}{\pi^{2(d+1)}}
\int dS dT f(Z+S,Y+S) g(Z-T, Y+T)
\exp -2 S^A_i T_A^i \,,
\ee
which is normalized so that $1*f =f*1= f$ and
gives rise to the  commutation relations
\be
\label{defr1}
[Y_i^A , Y_j^B ]_* = \gvep_{ij}\eta^{AB}\,,\qquad
[Z_i^A , Z_j^B ]_* = - \gvep_{ij}\eta^{AB}\,, \qquad
[Y_i^A , Z_j^B ]_* =0\,.
\ee
For $Z$-independent elements (\ref{star}) amounts to (\ref{wstar}).
The following  useful formulae hold true
\be
\label{uff1}
Y^A_i * = Y^A_i + \half\left ( \f{\p}{\p Y_A^i}-\f{\p}{\p Z_A^i}\right )
 \,,\qquad
Z^A_i * =
Z^A_i + \half\left ( \f{\p}{\p Y_A^i}-\f{\p}{\p Z_A^i}\right )
 \,,
\ee
\be
\label{uff2}
* Y^A_i  = Y^A_i - \half \left( \f{\p}{\p Y_A^i}+\f{\p}{\p Z_A^i}\right )\,,
\qquad
* Z^A_i  = Z^A_i + \half \left (\f{\p}{\p Y_A^i}+\f{\p}{\p Z_A^i}\right )\,.
\ee

An important property of the star product (\ref{star}) is that it
admits an inner Klein operator
\be
\label{iK}
{\cal K} = \exp {-2z_i y^i } \,,
\ee
where
\be
y_i = \f{1}{\sqrt{V^2}} V_B Y^B_i\,,\qquad
z_i = \f{1}{\sqrt{V^2}} V_B Z^B_i\,,
\ee
that obeys \cite{Vasiliev:2003ev}
\be
\label{Kf}
{\cal K} *f = \tilde{f}*  {\cal K}\,,\qquad {\cal K} *{\cal K}=1\,,
\ee
where  $\tilde{f}(Z,Y) = f(\tilde{Z} , \tilde{Y})$.

Following \cite{Vasiliev:2003ev} we introduce the fields $W(Z,Y;K|x)$, $B(Z,Y;K|x)$ and   $S(\theta,Z,Y;K|x)$,
where $B(Z,Y;K|x)$ is a zero-form while $W(Z,Y;K|x)$ and $S(Z,Y;K|x)$ are
connection one-forms in space-time and auxiliary $Z^A_i$ directions,
respectively
\be
W(Z,Y;K|x)=dx^\un W_\un (Z,Y;K|x)\,,\qquad
S(Z,Y;K|x)=\theta^A_i S^i_A (Z,Y;K|x)\,,
\ee
where $\theta^A_i$ is a short-hand notation for $dZ^A_i$.
Here we have introduced an outer Klein operator $K$ not introduced in \cite{Vasiliev:2003ev}, that obeys
\be
\label{K}
{K} *f = \tilde{f}*  {K}\,,\qquad { K} *{ K}=1\,,
\ee
with $\tilde f(\theta, Z, Y):= f(\tilde \theta, \tilde Z, \tilde Y)$ according to  (\ref{tila})
for all $o(d-1,2)$ vectors including $\theta^A_i$ (here is the difference with the inner Klein operator
$\K$, that does not affect $\theta^A_i$).
This construction is analogous to that of the $4d$ theory of \cite{more}. It is
equivalent to the setup of \cite{Vasiliev:2003ev} in the sector of dynamical HS fields, that obey
\be\label{K1}
W(Z,Y;-K|x) = W(Z,Y;K|x)\,,\q S(\theta,Z,Y;-K|x)= S(\theta,Z,Y;K|x)\,,
\ee
\be \label{K2}
B(Z,Y;-K|x)=-B(Z,Y;K|x)\,.
\ee
The fields of opposite $K$-parity not present in the construction of \cite{Vasiliev:2003ev} are topological
carrying at most a finite number of degrees of freedom each. Analogous fields
appear in the  $4d$ HS theory of \cite{more}.
They  can be interpreted as modules
playing an important role in the HS gauge theory \cite{Didenko:2015pjo}.
Therefore, we prefer to keep them  in the $d$-dimensional theory
as well, that is achieved via the $K$-dependence. To this end one has to relax
 conditions (\ref{K1}), (\ref{K2}).

All differentials anticommute with each other
\be
dx^\un dx^\um = - dx^\um dx^\un\,,\q
\theta^A_i \theta^B_j = - \theta^B_j \theta^A_i\,,\q
dx^\un \theta^B_j = - \theta^B_j dx^\un
\ee
but commute with all other variables except for  the Klein operator $K$ in the case of $\theta^A_i$.

As explained in Section \ref{lina},
the fields $\go$ and $C$, that now include both HS gauge fields and the topological ones, are identified with the ``initial data''
for the evolution in $Z$ variables,
\be
\go (Y;K|x) = W(0,Y;K|x)\,,\qquad C (Y;K|x) = B(0,Y;K|x)\,.
\ee
The $Z$ - connection $S$ is determined  in terms
of $B$ up to a gauge freedom.


The full nonlinear system of HS equations of \cite{Vasiliev:2003ev} is
\be
\label{WW}
\dr W+W*W =0\,,
\ee
\be
\label{DS}
\dr S+W*S+ S*W =0 \,,
\ee
\be
\label{DB}
\dr B+W*B-B*{W} =0\,,
\ee
\be
\label{SB}
S*B = B*{S}\,,
\ee
\be
\label{SS}
S*S = -\half( \theta^A_i \theta_A^i  + 4 g \Lambda^{-1}   \theta_i \theta^{i} B*K*\K )\,,
\ee
where  $\theta_i = \f{1}{\sqrt{V^2}} V_B \theta^B_i$. The field $B$ in this paper differs from that of \cite{Vasiliev:2003ev} by a factor of $K$: $B\to B*K$
while the coupling constant
$g\neq 0$ was  set to 1 in \cite{Vasiliev:2003ev} (it can be rescaled away by a field redefinition $B\to g^{-1} B$).

Condition (\ref{sp2}) admits a proper deformation
to the full nonlinear theory, \ie there exists a nonlinear  deformation
$\ta_{ij}$ of  $t_{ij}$,
\be
 [\ta_{ij}\,,  \ta_{nm}]_*=(\epsilon_{jn} \ta_{im} +i\leftrightarrow j)+ n\leftrightarrow m \,,
\ee
(see Section \ref{tnon}) allowing  to impose the conditions
\be
\label{tonlc}
D(\ta_{ij})=0\,,\qquad
 [S, \ta_{ij}]_*=0\,,\qquad [B, \ta_{ij}]_*=0\,,
\ee
which amount to the original conditions $[t_{ij}, \go]_* =[t_{ij}, C]_*=0 $
in the free field limit.

The system (\ref{WW})-(\ref{tonlc})
is invariant under the HS gauge transformations
\be
\label{gxz}
\delta \W = [\gvep , \W ]_* \,,\qquad \delta B = [\gvep \,, B]_* \,
\ee
with an arbitrary $\ta^{ij}$ invariant gauge parameter $\gvep$. It
is off-shell in the sense that it does not account for the
factorization of the ideal $I^{int}$ associated with the $sp(2)$ generators $\ta_{ij}$. In this paper, this factorization is reformulated in Section \ref{refine}
in the BRST language.

\section{BRST charge in the adjoint representation}
\label{brst}

It is convenient to formulate the $sp(2)$ invariance condition  and factorization
transformations in the BRST language. To explain the construction we start with
the general case. Let $T_\ga$ be generators of a Lie (super)algebra $\mathfrak{g}$, that
obey the (graded) commutation relations
\be
[T_\ga\,,T_\gb ]_\pm = f_{\ga\gb}^\gga T_\gga\,
\ee
with structure coefficients $f_{\ga\gb}^\gga$. The standard BRST operator is
\be
\label{Qgen}
Q:= c^\ga T_\ga - \half (-1)^{\pi(T_\gga)} f_{\ga\gb}^\gga c^\ga c^\gb b_\gga\,,
\ee
where the ghosts $c^\ga$ and $b_\gb$ obey graded commutation relations
\be
[c^\ga\,, b_\gb]_\pm = \delta^\ga_\gb\,,\q
[c^\ga\,,c^\gb]_\pm = 0\,,\q [b_\ga\,, b_\gb ]_\pm=0\,
\ee
at the condition that their $\mathbb{Z}_2$ grading is
\be
\pi( c^\ga) = \pi( b_\ga) = 1-\pi(T_\ga) \,.
\ee
So defined BRST operator $Q$ is nilpotent,
\be
\label{Q2}
Q^2=0\,.
\ee
When $Q$ acts on a left module $V$ generated from the vacuum $|0\rangle$ annihilated by $b_{\ga}$,
\be
b_{\ga}|0\rangle  =0\,,
\ee
 elements $|v\rangle \in V$ are $b_\ga$-independent,
\be
\label{vac}
\quad |v\rangle =v(c) |0\rangle\,.
\ee
The ghosts are endowed with the $\mathbb{Z}$ grading (ghost number)
\be
gh\, c^\ga = 1\q gh\, b_\ga =-1\,.
\ee
In  the lowest ghost degree, $Q$-invariance in the Fock module realization (\ref{vac}) implies $T_\ga$ invariance.

In application to HS theory we will use the adjoint action of $Q$ via
graded commutator in an appropriate associative algebra $A$ such that $Q\in A$
(assuming Weyl ordering of ghosts $c^\ga$ and $b_\ga$ in $A$),
\be
Q(a) := [Q \,, a]_\pm\,,\q \forall a\in A \,.
\ee
This setup has an important distinction from the left module realisation, mixing the $b_\ga$-dependent  and $b_\ga$-independent sectors in a  way most relevant to the HS problem. Namely, consider an element $\xi$ of the form
\be
\label{xi}
\xi=  \xi^\gb  b_\gb
\ee
with $c^\ga$, $b_\ga$--independent $\xi^\gb\in A$. The $c^\ga$\,, $b_\gb$-independent
sector of $Q(a)$ has the form
\be
\label{qx}
Q(\xi )\Big |_{b=0} := \half \{T_\gb \,, \xi^\gb\}_\pm   \,.
\ee
The transformation
 \be
 \label{qxi}
\delta a = Q(\xi)
\ee
for $\mathfrak{g} =sp(2)$  will be shown in Section \ref{refeq}
 just to describe the $sp(2)$ ideal factorisation  mentioned in the end of the previous section. On the other hand, the $Q$ invariance condition
 \be
Q(a) =0
\ee
in the sector linear in
 $c^\ga$ with $gh=1$
 implies invariance of $a$ under the adjoint action of $\mathfrak{g}$ modulo the ideal factorisation transformations (\ref{qxi}) because for
 \be
a= a_0 +  [a_1^\ga (c)\,, b_\ga]\q gh \, a =0
\ee
with $a_1^\ga(c)$ linear in $c^\gb$
\be
\label{QT}
[Q\,,a] = [Q\,,a_0 ] + \half \{T_\ga\,, a_1^\ga\}+\ldots\,,
\ee
where ellipses denotes some $b_\ga,c^\gb$-dependent terms (in the Weyl ordering).

 Note that in the left module realisation of the $Q$ complex
 with the vacuum (\ref{vac}) there is no room for the
 $b_{\ga}$--dependent terms as in (\ref{xi}) and, hence, transformations
(\ref{qxi}). By changing the vacuum conditions (\ref{vac}) one can replace
some of the invariance conditions by the factorisation transformations
(see, e.g., \cite{Buchbinder:2024gll} and references therein) but not to reach
both simultaneously as in the adjoint scheme.

On the other hand, if the associative algebra contains left and
right modules as, for instance, fermion modules in the fields (\ref{selW}), (\ref{selB}) of the SHS model of Section \ref{nonlsup}, then the first term in
(\ref{QT}) can be represented in the form of the second one. This implies that
in that case the naive invariance condition $[Q\,,a]=0$ is dynamically trivial reducing to identity by an appropriate gauge transformation with the gauge parameter $a_1$. Note that physical fields are $c^\ga$, $b_\gb$-independent while, as usual in the BRST approach, the gauge symmetry parameters in the gauge transformations are replaced by ghosts $c^\ga$.
As a result, in that case the extra components in the left and right modules ({\it e.g.}, $\gga$-traceful components of fermions in the supersymmetric model of Section \ref{nonlsup}) are only eliminated  by the factorisation
conditions (\ref{qx}). This
is just what doctor orders since, generally,
the invariance conditions and factorisation transformations remove the same components at the linear order but may, in principle, be in conflict beyond. Note, however, that this does not happen in the BRST free  version of the SHS model as can be shown with the aid of relations of Section \ref{osp}.
(For detail see Appendix B.)

Finally, let us make a few comments on possible reductions of $Q$ via
projectors. First of all, if the representation $T_\ga$ of $g$ is a direct sum of
several representations $T_{a\ga}$, instead of introducing different ghosts $c^{a\ga}$ to each of them  with $Q=\sum_a Q^a$ it is sometimes convenient
to introduce projectors $\Pi_a$, that obey
\be
\Pi_a \Pi_b = \Pi_b \Pi_a =\delta_{ab} \Pi_a\,.
\ee
This allows one to introduce unified notations
\be
T_\ga=\sum_a \Pi_a T^a_{\ga}\,,\q c_a = \Pi_a c \,.
\ee

Another application of the projectors most relevant to the
SHS theory analysis is to
 $\mathbb{Z}_2$ graded algebras and, in particular, superalgebras. Let $T_\ga =(T_{0\phi}\,,T_{1\phi'})$ obey
\be
[T_{0\phi}\,, T_{0\psi}]_\pm= f_{\phi\psi}^\rho T_{0\rho}\,,\q
[T_{0\phi}\,, T_{1\psi'}]_\pm = f_{\phi\psi'}^{\rho'} T_{1\rho'}\,,\q
[T_{1\phi'}\,, T_{1\psi'}]_\pm = f_{\phi'\psi'}^\rho T_{0\rho}\,.
\ee
In that case the BRST operator $Q$ (\ref{Qgen}) has the form
\be
\label{Z2Q}
Q= c^\phi T_{0\phi} + c^{\psi'} T_{1\psi'} - \half f_{\phi\psi}^\rho c^\phi c^\psi b_\rho + f_{\phi\psi'}^{\rho'} c^\phi c^{\psi'} b_{\rho'} -\half
f_{\phi'\psi'}^\rho c^{\phi'} c^{\psi'} b_\rho\,.
\ee
Clearly, discarding all generators and ghosts with primed indices reduces $Q$ (\ref{Z2Q}) to $Q_0$ associated with the even subalgebra with generators $T_{0\phi}$, that still obeys the nilpotency condition $Q_0^2=0$. One can formally
describe this situation with the help of projectors $\Pi_0$ and $\Pi_1$ that
project the ghosts $c^\ga$ and $b_\ga$ to the sectors 0 and 1, respectively.
In these terms dropping the terms with $\Pi_1 c$ and $\Pi_1 b$ does not violate
the nilpotency of $Q$. Clearly, one can proceed analogously for any (not necessarily
even) subalgebra of $g$ with the generators $T_0$. The superalgebra case is simply most relevant to the problem under consideration.

\section{Refinement of the old version}
\label{refine}

Here we describe a refined version of the $A$-model of \cite{Vasiliev:2003ev} that illustrates the idea from where the new coupling constants can come from.
The same time we reformulate the $sp(2)$-invariance and factorisation transformations in the BRST language, that significantly simplifies the whole setup.
It is this formulation, that will be extended to the SHS theory.

Namely, let
\be
\label{QBRST}
Q := c^{ij} \tau_{ij} - c^{i}{}_n c^{jn} b_{ij}\,,
\ee
where the nonzero anticommutation relations for the ghosts $c^{ij}$ and
the conjugated ghosts $b_{ij}$ are
\be
\label{cbij}
\{c^{ij}\,, b_{nm}\} = \delta^i_n \delta^j_m + \delta^i_m \delta^j_n \,.
\ee
By construction, the nilpotency condition (\ref{Q2}) holds true.

In these terms the $sp(2)$ invariance conditions (\ref{tonlc}) amount to
\be
\label{qsl2wb}
\{Q\,,\W\}_*=0\,,\q [Q\,, \B]_*= 0\,,
\ee
where $\W$, $\B$ now depend on the ghost variables $c^{ij}$ and $b_{ij}$.
 The original fields $W$, $B$ have the ghost number zero being, respectively, one- and zero- space-time
 ghost--independent differential forms. (The grading of differential forms is assumed to extend that of $c^{ij}$ and $b_{ij}$, \ie differential forms of odd degrees anticommute with $c^{ij}$ and $b_{ij}$.) The construction of $\ta_{ij}$ repeats that of  \cite{Vasiliev:2003ev} and will be explained in Section \ref{nonlsup} within its supersymmetric extension.

\subsection{Field equations}
\label{refeq}

The field equations acquire a most simple form in terms
of the BRST  extended connection
\be
\label{WQ}
\W = \dr_x +Q +W 
\ee
with the fields  $W$ and $\B$ below  depending on the  variables $\theta^A_i,Z^A_i,Y^A_i,K,c^{ij},b_{ij}$ and space-time coordinates $x$.

The non-linear on-shell system  takes the canonical form
\be
\label{Ww}
\W *\W =  \half( \theta_A^i \theta^A_i  + 4 g \Lambda^{-1}  \gga* F(\B) ),\qquad
\ee
\be\label{Wb}
[\W \,, \B]_* =0\,,
\ee
where
\be
F(\B) = \B + \sum_{n=2}^\infty g_n (c_2)*\underbrace{\B*\ldots* \B}_n
\ee
with some coefficients $g_n(c_2)$ that may depend on the $sl_2$ Casimir
operator $c_2$
\be
c_2:= \ta_{ij} *\ta^{ ij}\,.
\ee
(The $\B$-independent term is in principle also
possible but, contrary to the  $3d$ HS theory of \cite{PV},
where it  generates  the mass of the matter fields, its role
at $d>3$ is less clear.) Note that all
terms with nonzero $g_n$ at $n\geq 2$ are  most  likely ruled out by the locality conditions.

Also let us note that in the refined setup the $G_x +G_Z$ grading of the fields $\W$ and $\B$ is replaced by the grading
\be
G: = G_x +G_Z +G_c - G_b\,,
\ee
where $G_x$, $G_Z$, $G_c$ and $G_b$ count the degrees in $dx$, $\theta^A$, $c^{ij}$ and $b_{ij}$, respectively.

The central element  $\gga$ in (\ref{Ww}) is
\be
\label{Gu}
 \gga:= \theta^i \theta_{i} *K*\K \,.
\ee
That $\theta_A^i \theta^A_i$ is central is obvious. To see that $\gga$ is central one has to take into account that
$K*\K$ has zero star-commutator with everything except for $\theta^i$. As a result, potentially non-zero terms
in the commutator of $\gga$ with anything must contain $(\theta^i)^3$ which is zero since $\theta^i$ are anticommuting
while $i$ takes only two values (\ie as a three-form in a two-dimensional space).
The nontrivial component of the  curvature $\W*\W$ responsible for the nontrivial HS dynamics is $ \gga *\B$.

The system is formally consistent in the sense that the associativity
relations $\W*(\W*\W)=(\W*\W )*\W$ and $(\W*\W)*\B = \B* (\W*\W )$
equivalent to Bianchi identities
are respected by equations  (\ref{Ww}), (\ref{Wb}).

The parts of equations (\ref{Ww}), (\ref{Wb}) associated with $Q$ in
(\ref{WQ}) impose the $sp(2)$ invariance conditions (\ref{tonlc}) on the original
fields $W$ and $B$. Indeed, let
\be
\label{WBU}
\W = W + U_\W^{ij}b_{ij}+c^{nm} b_{ij} V_{\W nm}^{ij}+\ldots \,,\q \B = B +
U^{ij}_\B b_{ij}+c^{nm} b_{ij} V_{\B nm}^{ij}+\ldots \,,
\ee
where $W$, $B$, $U_{\W}$, $V_{\W nm}^{ij}$,  $U^{ij}_\B$ and $V_{\B nm}^{ij}$ are $c^{ij}$, $b_{ij}$--independent while $\ldots$ denote other $c$, $b$-dependent
terms.\footnote{Note that the fields $U_\W^{ij}$ and $U^{ij}_\B$ are space-time two- and one-forms still carrying, respectively, the  total gradings $G = 1$ and $0$ due to the factors of $b_{ij}$ in (\ref{WBU}).} Then the parts of equations
(\ref{Ww}), (\ref{Wb}), that are linear in $c^{ij}$ and $b_{ij}$-independent, imply $sp(2)$ invariance of $W$ and $B$  up to the terms in the ideal as
in (\ref{QT}).

On the other hand, the term $U^{ij}b_{ij}$ brings the $\tau_{ij}$--dependent
term to the \rhs of the equations,
\be
\label{dxW}
\dr_x W + W*W = -\half \{\tau_{ij} \,,U_W^{ij}\}_*  +\half( \theta_A^i \theta^A_i  + 4 g \Lambda^{-1}  \gga* G(B) )\,,
\ee
\be
\label{dxB}
\dr_x B + [W\,,B]_* = \half \{\tau_{ij} \,,U_B^{ij}\}_*\,,
\ee
which implies factorization of the field equations over the ideal of
elements proportional to $\tau_{ij}$.  Hence, as anticipated,  dynamical field
equations are concentrated in the $\tau_{ij}$-independent sector.

The system of equations (\ref{Ww}), (\ref{Wb})
is invariant under the HS gauge transformations
\be
\label{qgxz}
\delta \W = [\epsilon , \W ]_* \,,\qquad \delta \B = [\epsilon\,,  \B]_*
\,.
\ee
For  $\epsilon$ of the form
\be
\label{xib}
 \epsilon =\gvep   + \xi_\W^{ij} b_{ij}
\ee
with $c,b$-independent parameters $\gvep$ and $\xi_\W^{ij}$
the gauge transformations (\ref{qgxz}) reproduce
usual HS gauge transformations with the parameters $\gvep$ and the
factorization transformations with the parameters $\xi_\W^{ij}$, that
factor out terms proportional to $\tau_{ij}$ in $W$. Remarkably,
there is another gauge symmetry with the gauge parameters
$\xi_\B^{ij}$ responsible for the factorization in the $\B$-sector,
\be
\label{Bxi}
\delta \B = [\W\,, \xi_\B]_*\,,\q \delta \W = 2g\Lambda^{-1}\gga *\xi_\B\,.
\ee
Note that, in the original equations with $W$ and $B$, the first formula in
(\ref{Bxi}) makes no sense since $B$ was a zero-form while $W$ was a one-form, while in the BRST-extended equations these gauge transformations  with
\be
\label{bxi}
\xi_\B = \xi^{ij}_\B b_{ij}
\ee
have perfect sense due to the $Q$ term in  $\W$ (\ref{WQ}).
The factorization gauge transformations with the parameters
$ \xi^{ij}_\W$ and $ \xi^{ij}_\B$ put the system on shell.

In the BRST setup all fields are allowed to depend on the ghosts
$c^{ij}$ and $b_{ij}$. To control that the $sp(2)$ algebra, that
determines the field pattern of the theory, remains undeformed it is
necessary to guarantee that the original BRST charge $Q$ (\ref{QBRST})
as a part of the HS field $\W$ is not affected by the nonlinear corrections
of the theory. The same time, $\W$ can receive nontrivial dependence on the
ghosts in the other sectors, that may affect the form of
the field transformations but not their algebra.

\subsection{Example}
\label{exam}

To illustrate the idea consider a particular deformation of the equation (\ref{dxW}) with
 \be
 \label{ex}
  V_{\W} =2 g \Lambda^{-1} \gga*\G(c_2)*\B\,,
  \ee
 where
 \be
    \G(c_2) = \sum_{a=1}^\infty g_{a}\underbrace{c_2*\ldots * c_2}_a \,.
\ee
{Naively, the dependence on} $g_a$ {at } $a>0$ {is trivial as it can be
absorbed into a field redefinition}
\be
\label{FR}
\B\to \B'= \B* (1+  \G(c_2))^{-1}\,.
\ee
However, since by virtue of (\ref{twad}) derivatives in the $Y^A_i$ variables are
related to space-time derivatives, the resulting expression for $\B'$ in terms of
$\B$ is nonlocal if any of $g_a\neq 0$.
As such, the field redefinition  is not in the allowed locality preserving  class.

Alternatively, one can attempt to get rid of the terms with
$g_a\neq 0$  by the factorization gauge transformations (\ref{Bxi}) for $\B$.
Again, naively, all such terms can be represented in the form of gauge transformations (\ref{Bxi}) with some $\xi_\B$.
However, in all orders in $g_n$, $\xi_\B$ may not
 have the  projectively-compact spin-local form.

To make the system nontrivial on shell one has to specify the classes
of functions in which the gauge parameters $\gvep$ and, most important,
$\xi_{\W,\B}$ are valued. Generally, apart from the gauge transformation parameters
the factorisation transformations may depend non-linearly on the fields $W$ and
$B$ themselves. In this paper we do not consider their specific form
leaving detailed analysis of this problem for the future.

\subsection{Linearized analysis}
\label{lina}

The lowest-order analysis of the refined version of the
nonlinear HS equations is analogous to that of \cite{Vasiliev:2003ev}.
Indeed, let us set
\be
W=W_0 +W_1 \,,\qquad S= S_0 +S_1 \,,\qquad B=B_0 +B_1\,,\q U_{\W,B} = U_{0\W,B}+U_{1\W,B}
\ee
with the vacuum solution
\be
\label{BS0}
B_0 = 0 \,,\qquad S_0 = \theta^A_i Z_A^i \,,\qquad
W_0 =  \half \go_0^{AB} (x)  Y^i_A Y_{iB}\,,\q U_{0\W} = 0\,,\q U_{0\B}=0\,,
\ee
where $\go_0^{AB} (x)$ is demanded  obey the zero-curvature conditions
(\ref{Rvac}) to describe $(A)dS_d$. {}Then, it is not hard to see that equation (\ref{dxB})  in the $\theta^A_i$ sector together  with the factorization gauge transformations (\ref{bxi}) yield
\be
\label{BC}
B_1 = C(Y|x)\,
\ee
with some $sp(2)$ invariant traceless $C(Y|x)$.

Consider now equation (\ref{Ww}) in the $\theta^2$ sector
with $F(\B) =\B$. First of all we observe that
the gauge freedom (\ref{gxz}) (equivalently, (\ref{qgxz}) with the parameter
$\gvep$ in (\ref{xib})) allows us set all components of ${}^\pe S^A_{1i}$
to zero,
$$
S_1 = \theta_i s_1^i (z,Y|x) \,.
$$
The leftover gauge symmetry parameters are
${}^\pe Z$--independent. Equations (\ref{dxW}) and (\ref{dxB}) in the sectors of
Eqs.~(\ref{DS}) and (\ref{SB}) then demand
the fields $W$ and $B$ also  be ${}^\pe Z$--independent, \ie the
dependence on $Z$ enters only through $z_i$.
As a result, the $\theta^i \theta_i$ sector of (\ref{Ww}) amounts to
\be
\label{part}
\dr_z s_1
=-2 g \Lambda^{-1} \theta^i \theta_i
  C(-z ,{}^\pe Y)\exp{-2 z_k y^k} +\half \{\tau_{ij} \,,U_{1W}^{ij}\}_*\,.
\ee

Now we observe that $C(-z,{}^\pe Y)$ is spin-local-compact
since there is simply no room for $p_i^l p^{i n}$ with different $l$ and $n$ in the lowest order.
 As a result, equation (\ref{part}) takes the conventional form of \cite{Vasiliev:2003ev}
 in the  $U_{1\W}$-independent traceless sector.
The rest of the linearized analysis repeats
that of \cite{Vasiliev:2003ev} leading to the Central On-Mass-Shell theorem
(\ref{CMS1}), (\ref{CMS2}). Note that for spins $s\geq 1$ equation (\ref{CMS1}) expresses
$C(0,{}^\pe Y)$ via space-time derivatives of the dynamical (\ie
Fronsdal) HS gauge fields  contained in $\go(Y|x)$.

\subsection{Lessons}

Let us summarize the main lessons of the construction of the nonlinear $A$-type HS theory.

   The theory admits a set of operators $\ta_{ij}$ that form  $sp(2)$
  algebra in all orders in interactions as a consequence of the specific form of the
  HS equations. The  fields
   of the model are singled out by the two types of conditions, namely, that they are
   $sp(2)$ invariant
  \be
  \label{young}
D \ta_{ij} =0\,,\q \ta_{ij} *f = f*\ta_{ij}\,,
\ee
  and are equivalent up to the terms proportional to  $\ta_{ij}$,
  \be
  \label{trace}
f\sim f+ \ta_{ij}*g^{ij}\,, \q \ta_{ij} *g^{ij} = g^{ij}* \ta_{ij}\,.
\ee
The condition (\ref{young})
  singles out the fields described by traceful two-row Young diagrams of  $o(d-1,2)$  while
   (\ref{trace}) makes the fields $f$ and the HS gauge connections $W$ in $D= \dr + W $  traceless.

Both the $\tau_{ij}$ invariance conditions and $\tau_{ij}$ factorisation transformations admit a natural realization in terms of the BRST operator associated
with the $sp(2)$ generators $\tau_{ij}$.
Namely, the factorisation condition (\ref{trace}) is formulated in the form of
  gauge transformations (\ref{qgxz})-(\ref{Bxi}) associated with the new gauge fields $U$
 incorporated into the scheme as components of the fields $\W$ and $\B$, that depend
 on the conjugated $sp(2)$ ghost $b_{ij}$.  This is an essential
  modification  of the scheme allowing to control the functional class of the elements to be factored out, that, in turn, is anticipated to lead to a class of nontrivial vertices in the HS theory, that cannot be compensated by local
  field redefinitions.

Finally, the following comment is in order. The nontrivial part of the HS equations,
 namely (\ref{Ww}), can be generalized without violating  consistency to
\be
\label{SSR}
\W*\W = -\half( \theta^A_i \theta_A^i P_*(\B)   + \Lambda^{-1}   \theta_i \theta^{i}
R_*(\B)*K*\K )
\ee
with some
\be
P_*(c_2,\B) = \sum_{n=1,m=0}^\infty  p_{n,m} c_{2*}^m *\B_*^n \,,\q R_*(\B) = \sum_{n=1,m=0}^\infty  r_{n,m}c_{2*}^m *\B_*^n
\ee
with any star-product functions $P_*(c_2,\B)$ and  $R_*(c_2,\B)$.
In \cite{Vasiliev:2003ev} it was argued that such an extension is a sort of trivial because it can be eliminated by a field redefinition
\be
\label{BB'}
\B= R_* (c_2,\B')\,
\ee
and analogously for $\W$ and $P_*(\B)$.
(For more detail on this issue see the $4d$ paper \cite{more}.)
However, since such a field redefinition is nonlocal, not belonging
to the projectively-compact spin-local class ({\it cf.} the example of Section \ref{exam}), it may not be applicable,
\ie the higher-order  terms in $\B$ and $c_2$ cannot be compensated by a spin-local
field redefinition. We do not keep the terms nonlinear in $\B$ in
(\ref{Ww}) since they are anticipated to induce essentially nonlocal HS vertices
at the nonlinear level. Nevertheless, one should keep in mind that such terms
 can be easily reintroduced if necessary.

Now we are in a position to generalize the developed scheme to the
SHS theory  in any dimension, that unifies $A$ and $B$-models.

\section{Higher-spin supersymmetries in any dimension}
\label{SHSasg}

In this section we consider HS supersymmetries underlying the SHS theory that
describes both bosonic and fermionic HS fields and related generating functions
for the SHS multiplets.

\subsection{Oscillator algebra and the direct sum decomposition}
\label{osc}

The main idea of the  fermionic extension consists \cite{Vasiliev:2004cm}
of supplementing the
set of commuting variables $Y^A_i$ by the Clifford variables $\phi^A$,
that obey the relations
\be
\label{comre}
[Y^A_i\,,Y^B_j]_*= \gvep_{ij} \eta^{AB}\,,\q \{\phi_A\,,\phi_B\}_* =\eta_{AB}\,\q
(A=0,1,\ldots d\,,\q d=M+1)
\ee
with respect to the associative Weyl-Clifford star product
\bee
\label{sstar} f(Y,\phi)*g(Y,\phi) = &{}& \ls (2\pi)^{-2(M+2)} \int
d^{2(M+2)} S\, d^{2(M+2)} T\,d^{M+2}\ga\,  d^{M+2}\gb  \times
\nn\\
& \ls\ls\ls\ls&  \ls\ls\ls\ls\exp\big ( 2(\ga^A\gb_A- S^A_j T_A^j )\big ) f(Y+S,\phi+\ga)g(Y+T, \phi+\gb)\,
\eee
implying in particular the following useful relations
\be
\label{basr}
Y^A_i * = Y^A_i + \half \f{\overrightarrow{\p}}{\p Y^i_A}\,,\qquad
* Y^A_i  = Y^A_i - \half \f{\overleftarrow{\p}}{\p Y^i_A}\,,
\ee
\be
\label{basrc}
\phi^A * = \phi^A +\half \f{\overrightarrow{\p}}{\p \phi_A}\,,\qquad
*\phi^A =\phi^A +\half \f{\overleftarrow{\p}}{\p \phi_A}\,.
\ee

In terms of these oscillators we introduce the
$sp(2)$ generators
\be
\label{tij1}
t_{ij}:={Y}^A_i\, Y_{Aj}
\ee
and supergenerators
\be
\label{ti1}
 t_{i}:={Y}^A_i\phi_A\,,
\ee
that together form $osp(1,2)$,
\be
\label{osp12}
\ls \{t_i\,,t_j\}_* = t_{ij}\,,\q [t_{ij}\,,t_k ]_*= \gvep_{jk} t_i + \gvep_{ik} t_j\,,\q
[t_{ij}\,,t_{kl}]_* = \gvep_{jk} t_{il} + \gvep_{ik} t_{jl} +
\gvep_{jl} t_{ik} + \gvep_{il} t_{jk}\,.
\ee
They  rotate $osp(1|2)$ indices
$$
[t_{ij}\,,Y^A_k]_*=\gvep_{jk} Y^A_{i}+\gvep_{ik}Y^A_{j}\,,\q \{t_{i}\,,\phi^A\}_*=
Y^A_i\,,\q
[t_{i}\,,Y^A_j]_*= \gvep_{ij}\phi^A\,.
$$

  The ${o}(d-1,2)$ generators
\be T^{AB}={Y}^{Ai}{Y}^B_i + \phi^A\phi^B \,
\ee
{ rotate} $o(d-1,2)$ {indices}
\bee\nn
[T^{AB}\,,Y_i^C ]_*=Y_i^A\eta^{BC} -
Y_i^B\eta^{AC} \,,\q [T^{AB}\,,\phi^C ]_*=\phi^A\eta^{BC} -
\phi^B\eta^{AC} \,.
\eee
$T^{AB}$ {and} $t_{ij}$ {form a Howe dual pair},
$o(d-1,2)\oplus osp(1|2)$,
\be
[T^{AB},t_{ij}]_*=0\q [T^{AB},t_{i}]_*=0\,.
\ee

The oscillators $Y^A_i$ and $\phi^A$ can be unified into  superoscillators
\be
\Phi_\Omega^A = (Y^A_i, \phi^A )\,
\ee
with $\Omega = (i\,, \bullet)$, that obey the (anti)commutation relations
\be
[\Phi^A_\Omega\,,\Phi^B_\Lambda ]_{*\pm} = \eta^{AB} C_{\Omega \Lambda}\,,
\ee
where $C_{\Omega \Lambda}=(\gvep_{ij},\delta_{\bullet \bullet})$ is the $osp(1,2)$ invariant bilinear form and
$[\Phi^A_\Omega\,,\Phi^B_\Lambda ]_{*\pm}$ denotes the star-anticommutator at
$\Omega=\Lambda=\bullet$ and  star-commutator otherwise. In these terms the $osp(1,2)$
generators
\be
t_{\Omega\Lambda}= \Phi^A_\Omega \Phi^B_\Lambda \eta_{AB}\,
\ee
obey
\be
\label{tt}
[t_{\Lambda \Omega} \,, t_{\Phi \Psi} ]_* = C_{\Omega \Phi} t_{\Lambda \Psi} +
(-1)^{\pi_\Lambda \pi_\Omega}
 C_{\Lambda \Phi} t_{\Omega \Psi}+ (-1)^{\pi_\Psi\pi_\Phi} C_{\Omega \Psi} t_{\Lambda \Phi} +
(-1)^{\pi_\Lambda \pi_\Omega +\pi_\Psi \pi_\Phi} C_{\Lambda \Psi} t_{\Omega \Phi} \,.
\ee

Relations (\ref{tt})  hold true for $osp(m,n)$ with $\Omega = (  \alpha\,, i)$,
  $\alpha = 1,\ldots m$, $i=1,\ldots n$. At $m=1$, the supergenerators are
\be
t_i = t_{i\bullet}\,,
\ee
while the $o(n)$ generators $t_{\ga\gb}=-t_{\gb\ga}$ are absent. Note that, in the case of $osp(1,2)$, the latter fact
allows one to discard the sign factors like $(-1)^{\pi_\Psi\pi_\Phi} $ in (\ref{tt}).

To construct algebra $osp(1,2)$ that acts on solutions of the HS equations, we apply the following general scheme.
Let there be two $ osp(m,n)$ algebras $L$ and $T$ with the generators
$L_{\Omega\Lambda}$ and $T_{\Omega\Lambda}$ obeying the $osp(m,n)$
 relations,
\be
\label{ttl}
[L_{\Lambda \Omega} \,, L_{\Phi \Psi} ]_{\pm} = C_{\Omega \Phi} L_{\Lambda \Psi} +
(-1)^{\pi_\Lambda \pi_\Omega}
 C_{\Lambda \Phi} L_{\Omega \Psi}+ (-1)^{\pi_\Psi\pi_\Phi} C_{\Omega \Psi} L_{\Lambda \Phi} +
(-1)^{\pi_\Lambda \pi_\Omega +\pi_\Psi \pi_\Phi} C_{\Lambda \Psi} L_{\Omega \Phi} \,,
\ee
 \be
\label{ttt}
[T_{\Lambda \Omega} \,, T_{\Phi \Psi} ]_\pm = C_{\Omega \Phi} T_{\Lambda \Psi} +
(-1)^{\pi_\Lambda \pi_\Omega}
 C_{\Lambda \Phi} T_{\Omega \Psi}+ (-1)^{\pi_\Psi\pi_\Phi} C_{\Omega \Psi} T_{\Lambda \Phi} +
(-1)^{\pi_\Lambda \pi_\Omega +\pi_\Psi \pi_\Phi} C_{\Lambda \Psi} T_{\Omega \Phi} \,
\ee
and such that $T$ is in the adjoint representation of $L$,
\be
\label{ltt}
[L_{\Lambda \Omega} \,, T_{\Phi \Psi} ]_\pm = C_{\Omega \Phi} T_{\Lambda \Psi} +
(-1)^{\pi_\Lambda \pi_\Omega}
 C_{\Lambda \Phi} T_{\Omega \Psi}+ (-1)^{\pi_\Psi\pi_\Phi} C_{\Omega \Psi} T_{\Lambda \Phi} +
(-1)^{\pi_\Lambda \pi_\Omega +\pi_\Psi \pi_\Phi} C_{\Lambda \Psi} T_{\Omega \Phi} \,.
\ee
From these relations it follows that $T'$ with the generators
\be
T'_{\Lambda \Omega} := L_{\Lambda \Omega} - T_{\Lambda \Omega}
\ee
also forms $osp(m,n)$,
 \be
\label{ttt'}
[T'_{\Lambda \Omega} \,, T'_{\Phi \Psi} ]_\pm = C_{\Omega \Phi} T'_{\Lambda \Psi} +
(-1)^{\pi_\Lambda \pi_\Omega}
 C_{\Lambda \Phi} T'_{\Omega \Psi}+ (-1)^{\pi_\Psi\pi_\Phi} C_{\Omega \Psi}
 T'_{\Lambda \Phi} +
(-1)^{\pi_\Lambda \pi_\Omega +\pi_\Psi \pi_\Phi} C_{\Lambda \Psi} T'_{\Omega \Phi} \,,
\ee
that is in the adjoint representation of $L$. This construction will be used
in Section \ref{tnon} for the proof of the action of $osp(1,2)$ on the dynamical fields.

\subsection{U($osp(1,2))$ relations}
\label{osp}

Here we consider some relations obeyed by the $osp(1,2)$
generators, that result from the defining relations (\ref{osp12}) independently of
a particular representation.

Taking into account the second relation in (\ref{osp12})
one finds that $t_j$ obey the deformed oscillator commutation relations \cite{Aq}
\be
\label{ttQ}
[t_j\,,t_k]_* =  \half\epsilon_{jk} (1+q)
\ee
with some $q$ obeying
\be\label{tQ}
t_i *q = - q* t_i\,.
\ee
($q$ is called  odd Casimir operator.)
Equivalently,
\be
\label{Q}
t_i * t^i = \half (1+q)\,.
\ee
Other way around from {\ref{ttQ}), (\ref{tQ}) the $osp(1,2)$ relations
(\ref{osp12}) follow. This fact will be used in Section \ref{tnon} for the proof of $osp(1,2)$ associated with $s_i$.

Also, from (\ref{ttQ}) and (\ref{tQ}) it follows that
\be
\label{ttL}
t_{ij}*t^j = L* t_i \,,\qquad t^j *t_{ij} = - t_i*L \,
\ee
with
\be
\label{L}
L= \half (3 - q ) = 2- t_i *t^i \,.
\ee
One can  check that $L$ obeys
\be
\label{LL}
L*L=2L +\frac{1}{2} t_{ij} *t^{ij}\,.
\ee
Note that in \cite{Vasiliev:2004cm} $L$ was introduced in a specific
oscillator  realization of
$osp(1,2)$ (\ref{tij1}) and (\ref{ti1}) while here all relations are shown to hold true
 for any representation of $osp(1,2)$, that is these are relations in the universal
 enveloping algebra $U(osp(1,2))$. Formulae (\ref{ttL})-(\ref{LL}) are used in Appendix B.

\subsection{$B$ algebra}

Let $\cs$  be {the associative algebra} {of} $osp(1|2)$ {invariants}
\be
\label{YDB}
f(Y,\phi)\in \cs\,:\q [t_{ij}\,, f(Y,\phi)]_*=0\,,\q   t_{i} * f(Y,\phi) - f(Y, -\phi) * t_i =0\,.
\ee
These conditions imply  \cite{Vasiliev:2004cm}
that $f(Y,\phi)$ describe two-row hook-type tensors associated
with various degree-$2m$ polynomials in $Y^A_i$ and degree-$k$ polynomials in $\phi^A$,
\bee
\begin{picture}(45,50)
{
{\small \put(00,40){\line(1,0){40}}%
\put(00,35){\line(1,0){40}}%
\put(00,30){\line(1,0){40}}%
\put(00,25){\line(1,0){05}}%
\put(00,20){\line(1,0){05}}%
\put(00,15){\line(1,0){05}}%
\put(00,10){\line(1,0){05}}%
\put(00,05){\line(1,0){05}}%
\put(00,00){\line(1,0){05}}%
\put(00,00){\line(0,1){40}} \put(05,00.0){\line(0,1){40}}
\put(10,30.0){\line(0,1){10}}
\put(15,30.0){\line(0,1){10}} \put(20,30.0){\line(0,1){10}}
\put(25,30.0){\line(0,1){10}} \put(30,30.0){\line(0,1){10}}
\put(35,30.0){\line(0,1){10}} \put(40,30.){\line(0,1){10}}
}
\put(22,42.){\scriptsize  $m$}
\put(-30,20){\scriptsize  $k+2$}}
\end{picture}\,\q\q.
\label{kryuk}
\eee

{Algebra} $\cs$ {contains a two-sided ideal} {spanned by the elements}
\be
\label{fac}
g\in\ci :\qquad g=t_{\Phi\Lambda}* g^{\Phi\Lambda}  =g^{\Phi\Lambda} * t_{\Phi\Lambda}
\,,\q g\in \cs\,,
\ee
where
$g^{\Phi\Lambda}$ is in the adjoint representation of $osp(1,2)$,
\be
\label{tg}
[t_{\Lambda \Omega} \,, g_{\Phi \Psi} ]_* = C_{\Omega \Phi} g_{\Lambda \Psi} +
(-1)^{\pi_\Lambda \pi_\Omega} C_{\Lambda \Phi} g_{\Omega \Psi}+ (-1)^{\pi_\Phi \pi_\Psi}
C_{\Omega \Psi} g_{\Lambda \Phi} + (-1)^{\pi_\Lambda \pi_\Omega +{\pi_\Phi \pi_\Psi}}
 C_{\Lambda \Psi} g_{\Omega \Phi} \,.
\ee

{Type-$B$ HS algebra is} ${\mathcal B}:=\cs / \ci$.
 Its basis consists of  two-row hook traceless tensors of the form (\ref{kryuk})
 \cite{Vasiliev:2004cm}. Such fields are
associated with the mixed symmetry (hook-type) massless gauge bosons
 in $AdS_d$

 $$
\sum_{p,q }\oplus\,\,\,\,\,\,\,\,
{ \begin{picture}(80,20)(0,20)
{\linethickness{.250mm}
\put(00,50){\line(1,0){80}}%
\put(00,00){\line(1,0){10}}%
\put(00,10){\line(1,0){10}}%
\put(00,20){\line(1,0){10}}%
\put(00,30){\line(1,0){10}}%
\put(00,40){\line(1,0){80}}%
\put(00,0){\line(0,1){50}}%
\put(10,00.0){\line(0,1){50}} \put(20,40.0){\line(0,1){10}}
\put(30,40.0){\line(0,1){10}} \put(40,40.0){\line(0,1){10}}
\put(50,40.0){\line(0,1){10}} \put(60,40.0){\line(0,1){10}}
\put(70,40.0){\line(0,1){10}} \put(80,40.0){\line(0,1){10}} }
\put(22,55){ \small ${p+1}$} \put(-15,22){ \small ${q}$}
\end{picture}}$$ \\
except for the totally antisymmetric column-type fields, that are massive
\cite{Vasiliev:2004cm}.

The following comment is now in order. Naively one might think that
it is possible to remove traces between the $Y^A_i$ variables by factoring
 away elements proportional to $t_{ij}$ and those between $Y^A_i$ and $\phi^A$
by factoring away  elements proportional to $t_{i}$. This is not
the case because separately each of these factorizations is not $osp(1,2)$ invariant
violating the Young symmetry conditions (\ref{YDB}).
(A concomitant fact is that
the Young symmetry relates  traces in the two rows (\ie $Y^A_i$) to
those between the  column and one row, \ie $\phi^A$ and $Y^A_i$ variables.) The factorisation condition (\ref{fac}) on the other hand is $osp(1,2)$ invariant, eliminating traces in a way consistent with the Young properties of a tensor.

The $B$-algebra admits an element
\be \label{Gamma} \Gamma =(i)^{\half (M-2)(M-3)}
\phi^0\phi^1\ldots \phi^{M+1}\,,
 \ee
 that satisfies
 \be \label{Gcom}
\Gamma *\phi^A = (-1)^{M+1} \phi^A * \Gamma \,,\qquad \Gamma_*^2=Id\,
\ee and \be \label{gdag} \Gamma^\dagger =  \Gamma \,.
\ee
As a
result, the projectors
\be \label{pipm} \Pi_\pm := \half (1\pm
\Gamma )\,
\ee
are Hermitian,
\be \label{12} (\Pi_\pm )^\dagger
=\Pi_\pm\,.
\ee
According to (\ref{Gcom}), the elements $\Gamma$
and $\Pi_\pm$ are central for odd $M$ in which case one can use $\Pi_\pm$ to single out
the subalgebras $B_\pm$ of the $B$-algebra. On the other hand, one can consider the even subalgebra spanned
by even functions of $\phi^A$,
\be
B^E:\qquad f(Y,-\phi ) = f(Y,\phi)\,.
\ee
This algebra admits for any $M$
two subalgebras $B^E_\pm$ singled out by the projectors $\Pi_\pm$, that are central in $B^E$.

\subsection{Superalgebra}
\label{Higher Spin Superalgebra}


To define a superalgebra, that unifies $A$ and $B$ algebras, in
\cite{Vasiliev:2004cm} it was suggested to introduce two
sets of conjugated  spinor elements $\chi_\mu $ and $\bar{\chi}^\mu$, that
 commute with $Y^A_i$,
 \be
\chi_\mu  *Y_i^A = Y^A_i * \chi_\mu  \,,\qquad \bar{\chi}^\mu
* Y^A_i = Y^A_i *  \bar{\chi}^\mu\,
\ee
and form modules over the $o(M,2)$ Clifford algebra
($\mu , \nu\ldots$ are spinor indices),
\be \label{clre} \chi_\mu *
\phi^A= \gga^A{}_\mu{}^\nu \chi_\nu \,,\qquad \phi^A* \bar{\chi}^\mu
=  \bar{\chi}^\nu\gga^A{}_\nu{}^\mu \,,
\ee
where $\gga^A{}_\nu{}^\mu$ are  $o(M,2)$ gamma
matrices. Also in \cite{Vasiliev:2004cm} it was suggested to introduce two
projectors $\Pi_1$ and $\Pi_2$,
\be \Pi_1*\Pi_1 = \Pi_1\,,\qquad
\Pi_2*\Pi_2 = \Pi_2\,,\qquad \Pi_1 *\Pi_2 = \Pi_2*\Pi_1 = 0\,,\qquad
\Pi_1 +\Pi_2 = I\,,
\ee
demanding
\be \label{prtr} \Pi_1 *\chi_\mu =
\chi_\mu  * \Pi_2 =\chi_\mu \,, \qquad \Pi_2
*  \bar{\chi}^\mu =  \bar{\chi}^\mu* \Pi_1 =
\bar{\chi}^\mu\,, \ee
\be  \Pi_2 *\chi_\mu =
\chi_\mu  * \Pi_1 =0 \,, \qquad \Pi_1
*  \bar{\chi}^\mu =  \bar{\chi}^\mu* \Pi_2 =0\,,
\ee
\be \Pi_1 * \phi^A = \phi^A *\Pi_1 =
0\,,\qquad \Pi_2 * \phi^A = \phi^A *\Pi_2 =  \phi^A\,,\qquad
\{\phi^A ,\phi^B \}_* = \eta^{AB}\Pi_2\,.
\ee
As a result,
\be
\label{spinr21}
\phi^A*\chi_\mu =0\,,\qquad\bar \chi^\mu *\phi^A=0\,,\qquad   \bar{\chi}^\mu* \bar{\chi}^\nu=0  \,,\qquad \chi_\nu
* \chi_\mu \ =0\,. \ee

This projector
structure is conveniently  described by the auxiliary Clifford
variables obeying
\be
\Theta*\Theta = \bar{\Theta}*\bar{\Theta}=0 \,,\qquad \{\Theta , \bar{\Theta}\}_* = 1\,,
\ee
 that have zero (graded) star commutators with all
other generating elements. Then
\be
\label{p12}
 \Pi_1 := \Theta* \bar{\Theta}\,,\q
\Pi_2 :=  \bar{\Theta}*\Theta \,,
\ee
 $\chi_\nu $ contains one power of
${\Theta}$ and $ \bar{\chi}^\mu $ contains one power of
$\bar{\Theta}$. The following relations are useful
\be
\Theta * \Pi_2 *\bar\Theta = \Pi_1 \,,\q
\bar \Theta * \Pi_1 *\Theta = \Pi_2\,.
\ee
In addition,
 $ \chi_\nu * \bar{\chi}^\mu \in A \,,
 \bar{\chi}^\mu*\chi_\nu\in B \,,
 $
where the sectors of $A$- and $B$-algebras are associated with $\Pi_1$ and $\Pi_2$,
respectively.  (For detail see \cite{Vasiliev:2004cm}.)


In this paper we will use an equivalent formulation with the spinor part of
the SHS algebra  realised in terms
 of the Fock modules over the Clifford algebra of $\phi^A$ with the
 anticommutation relations (\ref{comre}).
  To this end, for even $M$, we decompose $\phi^A$ into creation and annihilation operators, $\phi^A= (\phi_+^{\A}\,,\phi_{-\A})$, that obey
\be
\{ \phi_+^{\A}\,, \phi_+^{\B}\}_* = 0\,,\qquad \{ \phi_{-{\A}}\,, \phi_{-{\B}}\}_* = 0\,,
\qquad \{ \phi_{-{\A}}\,, \phi_+^{{\B}}\}_* =\delta_\A^\B \,,\qquad \A,\B = 1,\ldots m = M/2\,.
\ee
For odd $M$ there is in addition a central element $\Gamma$ (\ref{Gamma}) that
obeys (\ref{Gcom}). In that case the generating  elements can be decomposed into
$(\phi_+^{\A}\,,\phi_{-\A}, \Gamma )$, $\A, \B = m=[M/2],$ where $[M/2]$ is
the integer part of $M/2$.

This allows one to introduce the  $\delta$-functions
\be
\label{delta}
\delta^{m}(\phi_-)* \delta(\bar \Theta)   :=\half (1+\Gamma_*^M)  \prod_{\A=1}^{m} \phi_-^1 \ldots \phi_-^{m}* \bar \Theta \,,\q
\delta(\Theta) *\delta^{m}(\phi_+) :=\half (1+\Gamma_*^M) * \Theta * \prod_{\A=1}^{m} \phi_{+1} \ldots \phi_{+m } \,,
\ee
that obey
\be
\phi_-^\A * \delta^{m}(\phi_-)*\delta({\bar \Theta})= 0\,,\q \delta^{m}(\phi_-)*\delta({\bar \Theta}) *\Pi_2 =0\,,
\ee
\be
\Pi_2 *\delta(\Theta)* \delta^{m}(\phi_+)= 0\,,\q \delta(\Theta)*\delta^{m}(\phi_+) *\phi_{+\A} =0\,,
\ee
\be
\Gamma * \delta^{m}(\phi_-)*\delta({\bar \Theta}) =  \delta^{m}(\phi_-)
*\delta({\bar \Theta})*\Gamma
= \delta^{m}(\phi_-)*\delta({\bar \Theta}) \,,
\ee
\be
\Gamma * \delta(\Theta)* \delta^{m}(\phi_+) = \delta(\Theta)*\delta^{m}(\phi_+)  *\Gamma = \delta(\Theta)* \delta^{m}(\phi_+)\,.
\ee

The Fock projector that obeys
\be
\phi_{- \A}* F=0\,,\q F* \phi_+^\A=0\,,\qquad F *F = F \,
\ee
can be introduced as
\be
F= (-1)^{\frac{m(m+1)}{2}} \Pi_1 * \,\delta^m (\phi_-)*\delta^m(\phi_+)\,.
\ee
At odd $M$, $F$ in addition obeys
 \be
 \Gamma * F =  F\,.
 \ee

In these terms spinor modules are realized as
\be
\label{chiphi}
\chi_\mu :  \quad \chi(\phi_+)  * \delta^m(\phi_-) * \bar \Theta \,,\qquad
\bar\chi^\mu : \quad  \Theta * \delta^m (\phi_+) * \bar\chi (\phi_-)\,,
\ee
where $\chi(\phi_+)$ and $\bar\chi(\phi_-)$ are arbitrary functions of $\phi_+$
and $\phi_-$, respectively.

The fields of the SHS theory can be represented  in the form
analogous to that of \cite{Vasiliev:2004cm}
\be
\label{sel}
a = a_{11} (Y)*F  +a_{22}(Y,\phi)*\Pi_2
 + a_{12}(Y,\phi_+)*\delta (\phi_-)* \bar \Theta +
 \Theta * \delta (\phi_+)* a_{21}(Y,\phi_-)  \,.
 \ee

 Elements  $a_{ij}(Y,\phi) $ are such polynomials of $Y^A_i$ and $\phi^A$ that they  commute with the $sp(2)$ generators $t_{ij}$ (\ref{tij1}) and $t_i$ (\ref{ti1})
 with the factor of $\Pi_2$ in the SHS theory,
\be
\label{ttp}
[t_{ij}\,, a]_*=0\,,\q [\Pi_2* t_i\,, a]_*=0\,.
\ee

The essential difference between the setup of this paper with that of \cite{Vasiliev:2004cm} is that now we do not insert the quasiprojector $\Delta$ to factor out all terms
proportional to $t_{ij}$ and $t_i$ since, as explained in Sections \ref{refine} and \ref{nonlsup}, in this paper such a factorization  is performed  by  gauging away  projectively-compact spin-local terms   proportional to $t_{ij}$ and $t_i$ .

The  fermionic fields $a_{12}$ and $a_{21}$ have to respect both
the (\ref{ttp}) invariance and  factorization conditions.
Because, in accordance with (\ref{sel}), $t_i$ contains a factor of $\Pi_2$ (\ref{p12}) while $a_{12}$ and
$ a_{21}$ are proportional to $\Theta$ and $\bar\Theta$, respectively, the former implies in particular
\be
\label{tistar}
t_i * a_{12} (Y,\phi_+)*\delta^m (\phi_-)=0\,,\q \delta^m (\phi_-)*a_{21} (Y,\phi_-) *t_i =0\,.
\ee
One can see that these conditions imply the $\gamma$-transversality of the respective
spinor-tensors, that makes them Lorentz irreducible. However, in addition one has to factor out the
terms of the form
\be
\label{tijk}
t_{\Lambda \Phi}*b^{\Lambda \Phi}_{12} \sim 0\,
\ee
provided that $t_{\Lambda \Phi}*b^{\Lambda \Phi}_{12}$ is $osp(1,2)$ invariant. Here is a potential subtlety: while the part of these conditions associated with
$t_{ij}$  implies the tracelessness of the spinor tensor in the $Y^A_i$
variables, the second one again contains
$\gamma$-traces. That is fine at the linearized level but  may be seemingly
problematic beyond if the latter condition  and (\ref{tistar})  are deformed
differently by the nonlinear corrections. In fact, this does not happen
because the expression (\ref{tijk}) must be $osp(1,2)$ invariant. As shown in Appendix B, this
can be seen with the aid of the relations of Section \ref{osp}.
However, by virtue of
the BRST technique presented in the next section, one can see that there is no problem whatsoever.

\section{Nonlinear supersymmetric equations}
\label{nonlsup}

Though the general idea of the construction of the SHS theory is analogous to
 that of the $A$-model,
some details are different and not completely obvious.

\subsection{Doubling of variables and the BRST charge}

Analogously to the type-$A$ HS theory  considered in Section \ref{A},
to formulate nonlinear field equations of the SHS theory we double
the variables $Y^A_i$ and $\phi^A$,
\be
Y^A_i \to (Z^A_i, Y^A_i)\,,\quad \phi^A \to (\psi^A, \phi^A)\,,
\ee
and introduce the star product
\bee
\label{sstar2} f(Y,\phi)*g(Y,\phi) = &{}& \ls (2\pi)^{-2(M+2)} \int
d^{2(M+2)} S d^{2(M+2)} Td^{M+2}\ga  d^{M+2}\gb
\exp( 2 ( \ga^A\gb_A-S^A_j T_A^j ) )\nn\\
& \ls\ls\ls\ls&  \ls\ls\ls\ls
f(Z+S,Y+S,\psi +\ga, \phi+\ga)g(Z-T, Y+T, \psi - \gb, \phi+\gb)\,,
\eee
that acts on both commuting $(Z^A_i,Y^A_i)$ and anticommuting $(\psi^A,\phi^A)$ variables.  One can see that  $(Z^A_i,\psi^A)$ have zero graded star
commutators with $(Y^A_i,\phi^A)$. (The integration variables $\ga$ and $\gb$ are
Grassmann odd.) The defining nonzero Weyl-Clifford commutation
relations are
\be
\label{basb}
[Y_i^A , Y_j^B ]_* = \gvep_{ij}\eta^{AB}\,,\qquad [Z_i^A , Z_j^B ]_* = -\gvep_{ij}\eta^{AB}\,,
\ee
\be
\label{basf}
\{\phi^A\,,\phi^B\}_* = \eta^{AB}
\,,\qquad  \{\psi^A\,,\psi^B\}_* = -  \eta^{AB} \,.
\ee

In addition, we introduce the differentials associated with the new variables
\be
\label{diff}
\theta^A_i = dZ^A_i\,,\quad \lambda^A=d\psi^A\,
\ee
at the convention that $\theta^A_i$ are odd while $\lambda^A$ are even so that
the differential
\be
\label{dthla}
\dr_{Z,\psi} := \dr_Z +\dr_\psi\,,\q \dr_Z:= \theta^i_A\f{\p}{\p Z_A^i}\,,\q
\dr_\psi:= \lambda^A\f{\p}{\p \psi^A}
\ee
is odd and nilpotent,
\be
\dr_{Z,\psi}^2 =0\,.
\ee

As in the $A$-model case, to control $osp(1,2)$ in the $B$-model it is most useful to use
the BRST formalism of Section \ref{brst}. The  BRST operator of the total
$osp(1,2)$ is
\be
Q:= c^{ij} \tau_{ij} + c^i \tau_i  - ( c^{i}{}_n c^{jn} + \f{1}{4} c^i c^j )b_{ij} -
2 c^{ij}c_i b_j\,
\ee
with the $sp(2)$ ghosts (\ref{cbij}) and the ghosts $c^i$ and $b_j$ associated with the
$osp(1,2)$ supergenerators $\tau_i$, that obey
\be
\label{cbi}
[c^i\,,b_j] = \delta_j^i\,.
\ee
So defined BRST charge obeys (\ref{Q2})
allowing to define the total differential
\be
\label{dq}
\dr:= \dr_{Z\psi}  +Q+\ldots \,,
\ee
where $\ldots$ denotes additional differentials associated with the homotopy coordinates, that appear in the
differential homotopy approach of \cite{Vasiliev:2023yzx}.  To guarantee nilpotency of  $\dr$  one has to demand
\be
\dr_{Z\psi} Q +Q\dr_{Z\psi} =0\,.
\ee
This condition is trivially obeyed at the linearized level, where $Q$ is $Z^A_i,\psi^A$--independent,
but, less trivially, as explained in Section \ref{tnon}, admits a nonlinear deformation, which property in fact determines the form of the HS equations.

With the collective variables
\be
\Y:=\{\theta^A_i,\lambda^A, Z^A_i,Y^A_i,\psi^A,K,c^{ij},b_{ij}\}
\ee
the  nonlinear equations in the SHS theory have the form in many respects
analogous to that of the $A$-model  (\ref{Ww})-(\ref{Wb}) with the fields
\be
\label{selW}
\ls\W =  \W_{11} (\Y)*\Pi_1*F +\W_{22}(\Y;\phi,c^i,b_i)*\Pi_2
 + \W_{12}(\Y;\phi_+,c^i,b_i)*\delta^m (\phi_-) *\bar \Theta  +
 \Theta * \delta^m (\phi_+)* \W_{21}(\Y;\phi_-,c^i,b_i)  \,,
 \ee
 \be
\label{selB}
\ls \B =  \B_{11} (\Y)*\Pi_1*F +\B_{22}(\Y;\phi,c^i,b_i)*\Pi_2
 + \B_{12}(\Y;\phi_+,c^i,b_i)*\delta^m (\phi_-) *\bar \Theta  +
 \Theta * \delta^m (\phi_+)* \B_{21}(\Y;\phi_-,c^i,b_i)  \,.
 \ee
Here fermions are described by the components $\W_{12}, \W_{21}, \B_{12}$ and
$ \B_{21}$ in which, in accordance with (\ref{chiphi}), spinor indices $\mu$ are associated with the space of functions of $\phi_+$ or $\phi_-$.

In the supersymmetric case the supergenerators of $osp(1,2)$ act on the fields of
the $B$-model and fermions but  not  on the fields of the $A$-model.
This property is expressed by the BRST operator
\be
\label{QS}
Q:= c^{ij} \tau_{ij} -  c^{i}{}_n c^{jn}b_{ij} +
\Pi_2 (c^i \tau_i  - \f{1}{4} c^i c^j b_{ij} -
2 c^{ij}c_i b_j)\,,
\ee
that still has the fundamental property (\ref{Q2}). Since $\Pi_2= \bar
\Theta \Theta$, the part of $Q$ (\ref{QS}), that depends on the super ghosts
$c^i$ and $b_i$  has the left action on  the fields
$\W_{12}$, $\B_{12}$, right action on $\W_{21}$, $\B_{21}$, adjoint action
on $\W_{22}$, $\B_{22}$ and trivial action on $\W_{11}$, $\B_{11}$. On the other hand, the $c^i$ and $b_i$--independent part of $Q$  associated with $sp(2)$ has adjoint action in all sectors.

\subsection{$osp^{tot}(1,2)$}
\label{osptot}

The generators $t^{tot}_{\Lambda \Omega}$ are by definition such that their even elements
$t^{tot}_{ij}\in sp(2)$  rotate $sp(2)$ indices $i$ and $j$ of all elements of the
algebra. This means that
\be
\label{Lt}
t^{tot}_{ij} := t_{ij}^{\theta} +t_{ij}^Z +t^{Y}_{ij}{}
\,,
\ee
where
\be
\label{Lth}
t_{ij}^\theta := \theta^A_i \frac{\p}{\p \theta^{Aj}} + \theta^A_j
\frac{\p}{\p \theta^{Ai}}\,,\q
t_{ij}^Z := - Z^A_i Z_{Aj}\,,\q
t_{ij}^Y :=  Y^A_i Y_{Aj}\,
\ee
with $t^Z$ and $t^Y$ acting via star-commutators, $t^{Z,Y} :=[t^{Z,Y},\,\, ]_*$, while $t^\theta$ acts as the differential operator (\ref{Lth}). Let us stress that
$t^\theta$ acts as an
outer operator of the algebra since $\frac{\p}{\p \theta}$ is not among its elements, \ie generating
functions in question depend on $\theta$ but on
$\f{\p}{\p \theta}$. However, being a vector field, $t_{ij}^\theta$ acts on the space of functions of $\theta$.

The next step consists of supersymmetrization of $t^\theta$, $t^Z$ and $t^Y$.
Let us start from $t^\theta$. Since $t^\theta$ acts on the anticommuting differentials $\theta$ the associated supergenerator has to involve the commuting superdifferentials $\lambda^A$ (\ref{diff}), $\lambda^A \lambda^B = \lambda^B \lambda^A $.
Setting
\be
\label{Qth}
t^\theta_i := (\lambda^A  \f{\p}{\p \theta^{A i}} + \theta^A_i \f{\p}{\p \lambda^A})\Pi_2\,
\ee
we observe that indeed
\be
\{t^\theta_i \,, t^\theta_j \} = t^\theta_{ij} \,.
\ee
The superpartners for $t^Y$ and $t^Z$ are, respectively,
\be
t^Y_i := Y^A_i \phi_A\Pi_2 \,,\q t^Z_i := Z^A_i\psi_A \,.
\ee
The total supercharge is
\be
\label{Qtot}
t^{tot}_i := t^\theta_i+ t^Y_i +t^Z_i\,.
\ee

The following comment is now in order. The supergenerator $t^\theta_i$ does not contribute
to the final result because the dynamical equations are concentrated in   the $\theta^i$, $\lambda$ - independent
sector, while $t^\theta_i$ (\ref{Qth}) brings the $\lambda$-dependent terms when acting on $\theta^i$ and does not contribute
when acting on the $\lambda$-independent terms. This allows us to drop the supercharge $t^\theta_i$ from our construction that, according to the general discussion at the end of Section \ref{brst}, does not affect the nilpotency of $Q$.  To see this more clearly we  consider the form of the nonlinear
SHS equations in some more detail.

\subsection{Nonlinear field equations}

To verify the $osp^{tot}(1,2)$
symmetry of the system we  check the compatibility of $Q$
(equivalently $t^\theta_{ij}$ and $t^\theta_i$) with the central element $\gga$ in (\ref{Ww}). That $t^\theta_{ij}$ commutes with
$\theta^{Ak} \theta_{Ak}$ in $\gga$  is obvious since the latter
is manifestly $sp(2)$ invariant. The situation with $t^\theta_i$ is less trivial since
\be
\label{Qtt}
t_i (\theta^{Ak}\theta_{Ak}) = 2 \lambda^A \theta_{Ai} \neq 0\,.
\ee

Naively, these terms can be compensated by the
replacement
\be
\label{C2tot}
\gga\to \tilde \gga := \delta^2 (\theta^\bullet_i)\delta(\lambda^\bullet)
  *K*\K \,.
\ee

(Note that $\theta^{\bullet i} \theta^{\bullet }_i =
\delta^2(\theta^\bullet_i)$.) Obviously, $\tilde \gga$
(\ref{C2tot}) is $t_i$ invariant,
\be
t_i (\tilde \gga) =0\,.
\ee
In fact, localized (integrable) functions of commuting differentials
known as integral forms  have been used in both math \cite{BL,Witten:2012bg} and physics
\cite{Zupnik:1989bw, Misuna:2013ww} literature.
However, now the problem is that $\delta^2 (\lambda^\bullet) *
\delta^2(\lambda^\bullet)$ diverges as $\delta^2(0)$ that makes the higher-order
corrections divergent in this setup. We were not able to find a manifestly $osp(1,2)$ invariant scheme free of this problem.
For that reason we leave $\gga$ in the original form
(\ref{Gu}) but modify the scheme in way, that preserves
$osp(1,2)$  action on the dynamical fields.

Namely, the field equations are postulated to keep the form (\ref{Ww}) and (\ref{Wb})
\be
\label{Wws}
\W *\W =  \half( \theta_A^i \theta^A_i  + 4 g \Lambda^{-1}  \gga* F(\B) ),\qquad
\ee
\be\label{Wbs}
[\W \,, \B]_* =0\,,
\ee
where all fields now depend on the additional variables according to (\ref{selW}),
(\ref{selB}). Since $\gga$ on the \rhs of (\ref{Wws}) is
$\psi\,, \lambda$-independent, hence
representing cohomology of $\dr_\psi = \lambda \f{\p}{\p \psi}$,
reconstruction of the perturbative corrections can also be
performed in the $\psi\,, \lambda$-independent way. As a result, $t_i^\theta$
(\ref{Qth}) acts trivially on the $\theta$, $\lambda$ independent physical
fields $\omega(Y,K)$ and $C(Y,K)$. Note that, naively, the same is true for
the $sp(2)$ generators $t^\theta_{ij}$, but in that case the manifest $sp(2)$
invariance in $\theta$ has to be controlled at every step of reconstruction
of the field equations in the physical sector  since the Poincar\'e
homotopy procedure involves the operator $\frac{\p}{\p \theta^i}$ that decreases
the power of $\theta$.

For
\be
\label{dQ}
\W:= \dr_Q+ \W'\,,\q \dr_Q := \dr_x +\dr
\ee
 with $\dr$ (\ref{dq})
 the field equations take the form
 \be
 \label{dwq}
\dr_Q\W' +\W' *\W'=2g\Lambda^{-1}
\gga * F(\B)\,,
\ee
\be
\label{dbq}
\D \B =0\,,\q \D := \dr_Q +[\W'\,,\,\,]_*\,.
\ee
The SHS field equations are invariant under the gauge transformations
(\ref{qgxz}), (\ref{Bxi}) with the appropriate modification of the fields
$\W$ and $\B$.

The BRST operator $Q$ has the form (\ref{QS}) with the generators $\ta_{ij}$, $\ta_i$, that we now define.

\subsection{Dynamical $osp(1,2)$}
\label{tnon}

 Derivation of the $sp(2)$ algebra and its $osp(1,2)$ extension  is based on the form of equation (\ref{Wws}) and is analogous to the $A$-model case \cite{Vasiliev:2003ev}.
 Namely, we observe that setting $Z_A^i \theta_i^A + S'= \theta_i^A S^i_A$ for
  the $S'$-component of $\W'$,  equation (\ref{Wws}) restricted on the $Q$--cohomology yields at $F(\B) = \B$
\be
\label{def}
\theta^A_i  S^i_A (K) *\theta^B_j  S^j_B (K)  \Big |_{H(Q)}  = -\half  (\theta^A_i \theta^i_A + 4 g \Lambda^{-1} \theta_i^\bullet \theta^{i\bullet}  *  K* \K *  \B)  \Big |_{H(Q)} \,.
\ee
The $g$-dependent  term on the \rhs of (\ref{def}) only depends on
$\theta_i^\bullet=\theta_i$. In the $\theta_i \theta^i$ sector (\ref{def}) yields
\be
\label{def1}
\theta_i  s^i (K) *\theta_j  s^j (K)  = -\half \theta_i \theta^i
  (1 + 4 g \Lambda^{-1}
  *  K* \K *  B) \,
\ee
(discarding in the sequel the cohomology restriction symbol $\big |_{H(Q)}$) with
\be
 s^i: =  \f{1}{\sqrt{V^2}}V^A S^i_A \,.
\ee

To move $\theta_i$ to the left on the \lhs of (\ref{def}) one has to take into account that $\theta_i$ anticommutes with $K$,
\be
\label{thetak}
\theta_i * K = - K* \theta_i\,.
\ee
By analogy with \cite{Vasiliev:2003ev} it is convenient to introduce an auxiliary
variable $\psi$ that obeys
\be
\label{rk}
\psi * K = -K* \psi\,,\q \psi* \theta_i = - \theta_i* \psi\,,\q
\psi*\psi = Id.
\ee
and the new field
\be
\hat s_i(\psi, K):= \psi* s_i(K)\,.
\ee
In these terms equation (\ref{def}) acquires the form
\be
\hat s_i* \hat s^i =  1 + 4 g \Lambda^{-1}
  *  K* \K * \B \,
\ee
of the
deformed oscillator algebra (\ref{ttQ}) \cite{Aq}
originally found by Wigner in \cite{wig} in a particular representation.
Indeed, thanks to the first relation in (\ref{rk}) $K* \K * \B$ anticommutes with $\hat s_i$.
As a result $\hat s_i$ behave as the generators $t_i$ (\ref{ttQ})  allowing to construct the $osp(1,2)$ generators
\be
t_{ij}:= \{\hat s_i\,, \hat s_j\}_*\,,\q t_i :=\hat s_i \,.
\ee
Using again (\ref{rk}) we find the $\psi$-independent expression for $t_{ij}$
\be
t_{ij}:= s_i (-K)*   s_j (K) + s_j (-K)*   s_i (K)\,.
\ee

Now we introduce
\be
\label{T}
\T_i:= \psi_A S^A_i\,,\q \T_{ij}:= \{\T_i\,,\T_j\}_*\,.
\ee

{} Taking into account that the $V$-transversal  in $\psi^A$ components of
$\T_i$ and $\T_{ij}$  form $osp(1,2)$
as in the case of undeformed oscillator algebra considered in Section
\ref{osc}, from the above analysis with
\be
\psi:= \f{1}{\sqrt{V^2}}V^A \psi_A
\ee
it follows,  that $\T_i$ and $\T_{ij}$ (\ref{T}) form $osp(1,2)$.

 In any $sp(2)$ covariant gauge in which $S_i^A$
is expressed in terms of $\B$ with no external $sp(2)$
noninvariant parameters (see \cite{Vasiliev:2003ev}),
$sp^{tot}(2)$ acts covariantly on $S^i_A$,
\be
[t^{tot}_{ij} \,, S^A_n ]_* =  \gvep_{in}S^A_{j}+ \gvep_{jn}S^A_{i}
\,.
\ee
Note that, analogously to the $A$-model case,  here we disregarded the term $\f{\p S^A}{\p  C} [t^{tot}_{ij} \,, C ]_*$
 accounting for  the $sp(2)$ transformation of the fields $C$ on which
$S^A_i$ does depend (see Eq.~(\ref{BC1}) of  Section \ref{pertshs}) because, by construction,  $C$ is demanded to be $sp(2)$ invariant.

As a result,  the operators
\be
\ta_{ij}:=t^{tot}_{ij}-\T_{ij}\,\q
\ta_i : = (t^{tot}_i - \T_i)\Pi_2\,
\ee
form  $sp(2)$ or $osp(1,2)$ in the $\Pi_2$ sector
 and commute with $S^A_i$, which means that nonlinear corrections
due to the evolution along $Z$-directions
do not affect the $sp(2)$ and $osp(1,2)$ algebras. The $Q$-invariance conditions resulting from
(\ref{dwq}) and (\ref{dbq}) imply  usual $sp(2)$ and $osp(1,2)$ invariance conditions (modulo terms in the ideal)
for the fields $\W$ and $\B$ and are identically satisfied
 on $S_i^A$.
In the free field limit with $S^A_i = Z^A_i$, $\ta_{ij}$
and $\ta_i$ coincide with (\ref{tij1}) and (\ref{ti1}),
respectively. It is important to stress that the $osp(1,2)$ generators
$t^\theta_{ij}$ and $t^\theta_i$ do not act on the $\theta$, $\lambda$-independent
terms. This allows us to neglect the contribution of $t_i^\theta$ at all
stages of the computation which is crucially important in the SHS model
where $\gga$ is not $\ta_i^\theta$ invariant.

\subsection{Perturbative analysis}
\label{pertshs}

The perturbative analysis of equations (\ref{Wws}), (\ref{Wbs})
repeats in main features  that of Section \ref{lina}.
The difference  is that the differential
$\dr_x+\dr_Z$  is replaced by $\dr_Q$ (\ref{dQ}), (\ref{dq}).

 We set
\be
\W'=\W'_0 +\W'_1 \,,\qquad \B=\B_0 +\B_1
\ee
with the vacuum solution
\be
\label{BS02}
\B_0 = 0 \,,\qquad
\W'_0 =  \half \go_0^{AB} (x) ( Y^i_A Y_{iB} + \phi_A \phi_B)\,,
\ee
where $\go_0^{AB} (x)$ satisfies the zero curvature conditions (\ref{Rvac}) to describe $(A)dS_d$. Here $S_0$ is set to zero since the effect of the more traditional
expression $S_0 = \theta^A_i Z_A^i$ is accounted by  $\dr_{xZ\psi}=\dr_x +\dr_{Z,\psi}$ in $\W$.
 {}From (\ref{Wbs}) and (\ref{BS02}) we obtain
\be
\label{BC1}
\B_1 = C(Y,\phi,\Theta,\bar\Theta|x)* K + U^{ij}(Y,\phi,\Theta,\bar\Theta|x)* K b_{ij}+\ldots\,.
\ee
Note that all terms with nonzero powers of $\ta_{\Phi\Psi} * \tau^{\Phi\Psi}$ and, hence, $t_{\Phi\Psi} * t^{\Phi\Psi}$ at the linearized
level can be gauged away by virtue of the factorization transformations (\ref{Bxi}).
Equation (\ref{Wbs}) demands the field $C(Y,\phi|x)$ to obey the twisted covariant constancy equation along with the $sp(2)$ invariance condition extended to $osp(1,2)$ beyond the $A$-sector.

Now consider  equation (\ref{Wws}) in the $\theta^2$ sector.
 First of all we observe that, using the gauge freedom (\ref{qgxz}),
 we can gauge away all components of ${}^\pe S^A_{1i}$, that allows us
 to set
\be
S_1 = \theta_i s_1^i (z,\psi,Y,\phi,\Theta,\bar\Theta|x) \,.
\ee
The leftover gauge symmetry parameters are  ${}^\pe Z$--independent.
Equations (\ref{Wws}) and (\ref{Wbs}) then demand
 $W$ and $B$ also  be ${}^\pe Z$--independent.
So, the dependence on $Z$ now enters only through $z_i$.
As a result, (\ref{Wws}) amounts to
\be
\dr_{Z\psi} s_1 =   2 g \Lambda^{-1}  \gga *  B\,
\ee
with $\gga$ (\ref{Gu}).
With the help of the relation
\be
\label{fK}
f(z,y)* \K = f(-y,-z)\exp (-2 z_k y^k )
\ee
for $\K$ (\ref{iK}) one obtains in the first order
\be
\label{partS}
\dr_{Z\psi} s_1 =2  g \Lambda^{-1}
\theta_i \theta^i C(-z,{}^\pe Y,\phi,\Theta,\bar\Theta)\exp (-2 z_k y^k )\,.
\ee

We observe that the $\rhs$ of (\ref{partS}) is projectively-compact spin-local
since it is linear in $C$ and, hence, the spin of the $l.h.s.$ is the same as  of $C$, that implies compactness.
The number of derivatives in the
linearized field equations  is limited because the background $AdS$ connection carries at most two space-time indices $A,B$.
This implies spin-locality.

That the $\psi$, $\lambda$--independent factor of $\gga$
represents cohomology of the differential $\dr_\psi$ (\ref{dthla})
allows us to keep this factor intact during the
analysis of the HS field equations.
As a result, we obtain
\be
\label{S1}
s_1 =\theta^j \f{\partial}{\p z^j}   \varepsilon_1 +2g\Lambda^{-1} \theta^j z_j
\int^1_0dt\,t
C(-t z,{}^\pe Y,  \phi,\Theta,\bar\Theta )\exp (-2 t z_k y^k )\,.
\ee

The freedom in the function $ \varepsilon_1=\varepsilon_1 (Z,\psi,Y,\phi,\Theta,\bar\Theta|x) $
manifests invariance under the gauge transformations
(\ref{qgxz}).
In the lowest order
it is convenient to fix an  $sp(2)$ invariant gauge by demanding $ \partial_i
\varepsilon_1=0 $.
The leftover gauge transformations  with $Z,\lambda,\psi$-independent parameters
 \be
\label{WE}
\varepsilon_1 (Z,\psi,Y,\phi,\lambda,\Theta,\bar\Theta|x)=\varepsilon_1 (Y,\phi,\Theta,\bar\Theta|x)\,
\ee
identify with the HS gauge transformations
acting on the $Z,\lambda,\psi$-independent  dynamical HS fields.
 As a result, the field $ S $ is
expressed in terms of $ C $.

The next step is to analyze linearized
equation (\ref{Wws})  in the $\theta dx $ sector,
\be
\label{partw}
\dr_{Z,\psi} W_{1x} = \dr_x s_1 +W_0 * s_1 +  s_1  * W_0  \,.
\ee
This yields by the standard Poincar\'e homotopy formula
\be
\label{W1}
W_{1x}(Z,Y)=\go (Y) -  2 g\Lambda^{-1}  z^j
 \int^1_0 dt\,(1-t)e^{-2tz_i y^i}
   E^B\f{\p}{\p Y^{jB}} C(-tz, {}^\pe Y,\phi )\,
\ee
(note that the terms with $ z_i \dr_x s_1^i$
  vanish because $ z^i z_i  = 0 $). Since, perturbatively,
the system as a whole is  a consistent
 system of differential equations with respect to the total differential
 $ \dr_Q$,
it suffices to analyze the $dx dx$ sector of
 (\ref{Wws}) and $dx$ sector of (\ref{Wbs}) at $ Z =0 $.
Thus, to derive dynamical HS equations,
it remains to plug (\ref{W1}) into (\ref{Wws}) and (\ref{BC1})
into (\ref{Wbs}), interpreting $ \go (Y,\phi|x)$ and
$ C(Y,\phi|x)$  as HS generating functions.

The elementary analysis  with the help of (\ref{uff1}) and (\ref{uff2})
yields
\be
\label{CMS1s}
R_1 ({}^\pa Y , {}^\pe Y,{}^\pa \phi , {}^\pe \phi,\Theta,\bar\Theta |x ) = \half g \Lambda^{-1}
E_0^A   E_0^B
\f{\p^2}{\p Y^A_i \p Y^B_j} \gvep_{ij} C(0,{}^\pe Y,  \phi,\Theta,\bar\Theta|x)\,.
\ee
For $B=C * K$, equation  (\ref{Wbs})
amounts to  (\ref{CMS2}).
Thus it is shown that the linearized part of the HS equations
(\ref{Wws}), (\ref{Wbs}) yields the generalization of the Central
On-Mass-Shell theorem for the SHS theory, that reproduces that of the
$A$-model, describes
fermions on the sector odd in $\Theta$ and $\bar \Theta$ and agrees with the result
presented in \cite{Grigoriev:2018wrx} for the $B$-model. In \cite{Grigoriev:2018wrx}
 it was also argued that a $B$-model has to exist at the nonlinear level. However, while the
 system (\ref{Wws}), (\ref{Wbs}) allows one to
systematically derive  all higher-order corrections to the free
equations as well to explore their locality properties, the formal approach of \cite{Grigoriev:2018wrx} is hard if at all possible to implement beyond the linearized approximation.

\subsection{Inner symmetries and truncations}
\label{inn}

The proposed system of
gauge invariant  nonlinear dynamical equations  for a SHS theory in $AdS_d$, that unifies $A$ and $B$ bosonic HS models with massless fermions, admits a   generalization  to SHS models with
unitary, symplectic and orthogonal  Yang-Mills groups. This is because, analogously to the $d=4$ case  \cite{Ann,KV}, the system (\ref{Wws}), (\ref{Wbs}) is consistent with the  fields
$\W $, and $B$ valued in any associative algebra $A$ thus describing an $A_\infty$ strong homotopy
algebra \cite{stash}. In particular, one can choose $A=Mat_p(\mathbb{C})$.
(Note that  the unfolded form of the pure Yang-Mills theory has been recently
worked out in \cite{Misuna:2024dlx}.)

For the SHS theory it is appropriate to use a $Z_2$-graded $A$. For instance, one can chose
$A=Mat_{n+m}(\mathbb{C})$ with even elements $a_{i'}{}^{j'}$ and $a_{i''}{}^{j''}$,  $i',j' = 1,\ldots n$,
$i'',g'' = 1\ldots m$ and odd ones $a_{i'}{}^{j''}$ and $a_{i''}{}^{j'}$.   In these terms, the fields (\ref{selW}), (\ref{selB}) acquire the form
\bee
\label{selp}
 \Phi =&&\ls  \Phi_{i'}{}^{j'} (\theta;Z;Y; K)* F  +
 \Phi_{i''}{}^{j''} (\theta,Z;Y;\lambda;\psi,\phi, K)\nn\\
 &&
\ls\ls + \Phi_{i'}{}^{j''}(\theta;Z;Y;\lambda;\psi;\phi_+;K)* \delta^m (\phi_-)  +
 \delta^m (\phi_+) *\Phi_{i''}{}^{j'}(\theta; Z;Y;\lambda;\psi;\phi_-;K)  \,
 \eee
for $\Phi=\W$ or $\B$. Here even and odd fields are, respectively, space-time tensors and
spinors. The respective SHS algebras with the pseudoorthogonal algebra $o(p,q)$
will be denoted $hgl(n,m|sp(2)[p,q])$. With these notations the algebra considered in the previous sections is $hgl(1,1|sp(2)[p,q])$.

The reality conditions
\be
\W^\dagger_i{}^j{}(Z,\psi,Y,\phi,K|x)= - (i)^{\pi(\W)} \W_i{}^j{} (Z, \psi,Y,\phi,K|x)\,,
\ee
\be
B^\dagger_i{}^j{} (Z,\psi,Y,\phi,K |x)= - (i)^{\pi(B)} {B}_i{}^j{} (Z,\psi,Y,\phi,K|x)
\ee
 give rise to a system with the global HS symmetry algebra
$hu(n,m|sp(2)[p,q])$ at the conditions
\be
(Y^A_i)^\dagger = Y^A_i\,,\q (Z^A_i)^\dagger = Z^A_i\,,\q \psi^{A\dagger} = \psi^A
\,,\q \phi^{A\dagger} = \phi^A \,,\ee
\be K^\dagger = K \,,\q (\theta^A_i)^\dagger =
\theta^A_i\,,\q (\lambda^A)^\dagger = \lambda^A\,,
\ee
that reduces the action of $\dagger$ to the reordering of the
product factors. Note that we use notations $i=(i',i'')$ with $\pi(A_{i'}{}^{j'})
= \pi(A_{i''}{}^{j''})=0$, $\pi(A_{i'}{}^{j''})
= \pi(A_{i''}{}^{j'})=1$. All fields in this algebra, including the spin-one
fields, that correspond  to
the $Z,\psi,Y,\phi$-independent part of $W_i{}^j(Z,\psi,Y,\phi|x)$, are valued in
$u(n)\oplus u(m)$ which is the Yang-Mills algebra of the theory.

Combining the antiautomorphism of the star product algebra
$\rho (f(Z,Y)) = f (-iZ,\psi,iY,\phi)$ with some antiautomorphism of the matrix
algebra generated by a nondegenerate form $\rho_{ij}$
one can impose the conditions
\be
\W_i{}^j  (Z,\psi,Y,\phi|x)=
- \rho^{jl}\rho_{k i} \W_l{}^k (-iZ,\psi,iY,\phi|x)\,,
\ee
\be
B_i{}^j  (Z,\psi,Y,\phi|x)=
- \rho^{jl}\rho_{ki} {B}_l{}^k (-iZ,\psi,iY,\phi|x)\,,
\ee
which
truncate the original system to the one with the Yang-Mills
gauge group $USp(n)\times USp(m) $ or $O(n)\times O(m) $
depending on whether the form
$\rho_{ij}$ is antisymmetric or symmetric, respectively (for more detail
see \cite{KV} or review \cite{Vasiliev:1999ba} for the $4d$ example).
The corresponding global HS symmetry algebras are called
$husp(n,m|sp(2)[p,q])$ and $ho(n,m|sp(2)[p,q])$,
respectively.
In these cases  all fields of odd spins are in the
adjoint representation of the
Yang-Mills group while the fields of even spins are in the opposite
symmetry second rank representation (\ie symmetric for
 $O(n)\times O(m)$ and antisymmetric for
 $USp(n)\times USp(m)$) which contains a singlet. The genuine graviton
is always the singlet spin two particle in the theory.
Color spin two particles are also included for general $n,m$,
however.\footnote{Note that this does not
contradict to the no-go results of \cite{CW,ng2} because
the theory under consideration does not allow a flat limit with  unbroken
HS  and color spin two symmetries.} The minimal HS  $A$-model of \cite{Vasiliev:2003ev} is  based on the algebra $ho(1,0|sp(2)[d-1,2])$.
It describes  even spin particles, each in one copy. Odd spins
do not appear because the adjoint representation of $o(1)$ is trivial.
Its generalization to the $B$-model  is  associated with the algebra $ho(0,1|sp(2)[d-1,2])$.

Also note that the chiral superalgebras $\supm$ result from $\su$
with the aid of the projectors $\Pi_\pm$ (\ref{pipm})
\be
\label{chel}
f\in \supm : \quad f=\Pi_\pm *g *\Pi_\pm \,,\quad g\in\su\,.
\ee
For even $M$ the projection (\ref{chel}) implies chiral projection
for spinor generating elements and projects out bosonic elements
 odd in $\phi$. For odd $M$ it implies irreducibility of the spinor
 representation of the Clifford algebra.

\section{Discussion}
\label{disc}

In  this paper  a new  class of  HS gauge theories in any dimension,
that contain both bosons and fermions and are invariant under HS supersymmetries,
is presented.
In the most cases (\ie for sufficiently large space-time
dimension $d$) the proposed theories are not supersymmetric in the usual sense since
anticommutators of the lower-spin supersymmetry generators contain HS generators in addition to the usual space-time ones. (This is simply because the symmetrized product
of the appropriate spinor representations contains a bunch of bosonic generators
in addition to the usual translation and Lorents generators associated with the
$\gga$-matrices $\gga_{AB}$ in this setup.)

  The proposed models, including the bosonic $A$-- and $B$--models as well as their
  supersymmetrization,  are conjectured to possess an infinite number of coupling constants associated with
the independent vertices found by Metsaev in \cite{Metsaev:2005ar}.
Naively, one might think that such coupling constants may be related by the
infinite-dimensional HS symmetry, but this is unlikely the case as follows from
the fact confirmed by the analysis of \cite{Tatarenko:2024csa}, that, relating
different spins, HS symmetries do not relate vertices with the same spins but different maximal numbers of space-time derivatives. For $d>4$ this leads to
infinite towers of vertices with two independent coupling constants.

The construction of this paper possesses a number of new elements, both technical and conceptual.

Technically, there are several important points. One is that the differentials
$\lambda^A= d \psi^A$ for the superpartners $\psi^A$ of $Z^A_i$ are commuting.
Together with the anticommuting differentials $\theta^A_i=dZ^A_i$ these form
an $osp(1,2)$ multiplet $(\theta_i^A, \lambda^A)$. In the construction of the
$A$-model, the cohomological term, that induces nontrivial interactions via
(\ref{Ww}), has  the form
\be
\label{deldel}
\delta^2 (\theta_i) \delta^2 (z_i)* \delta^2 (y_i) \,,
\ee
where the factor of $\delta^2 (y_i)$ regularizes this expression via
\be
\delta^2 (z_i)* \delta^2 (y^i) =\exp{-2 z_i y^{i}}\,.
\ee

In presence of $\psi$ and $\lambda$ the naive extension of the expression (\ref{deldel})
\be
\delta^2 (z_i)* \delta^2 (y_i) \delta(\lambda^\bullet) \delta (\psi^\bullet)\,
\ee
does not work since it develops the $\delta(0)$-type divergencies at the nonlinear
level.
This forced us to replace the manifestly $osp(1,2)$ invariant  expression
by the usual one that only respects manifest $sp(2)$ symmetry. Remarkably, this
is still compatible with the all order $osp(1,2)$ invariance on the dynamical
fields which is necessary for the interpretation of the theory in terms of the
dynamical fields that have to obey the $osp(1,2)$ invariance and factorisation
conditions.

Second point is that the formalism proposed in this paper operates with the
$osp(1,2)$ symmetry of the model in terms of the associated BRST operator,
that greatly simplifies the formulation, allowing  to pack in the same HS equations both $osp(1,2)$ invariance and factorisation conditions by virtue of introducing
respective ghosts as additional variables on which all HS generating functions depend.
It is important that this formulation admits a natural extension to the differential
homotopy approach of \cite{Vasiliev:2023yzx}.
 An interesting feature of the proposed BRST formalism based on the adjoint action of the BRST operator on the fields is that it automatically puts the system
on shell, generating the factorisation transformations, that remove the off-shell degrees of freedom. It would be interesting to work out a version of the BRST formalism
appropriate for the description of the off-shell HS theory.

The conceptual point is that to identify independent coupling constants in the theory it is necessary to
restrict the class of field redefinitions. As argued in
\cite{Vasiliev:2022med} the appropriate class is of the
so-called projectively-compact spin-local functions. Spin locality implies that the field redefinition is
local for any finite subset of fields involved into the field redefinition. Projective compactness implies
both that the field redefinition still makes sense for the  infinite towers of fields (this is referred as
compactness) and that spin-locality
takes place  both in terms of auxiliary variables like $Y$ and $\phi$ and in terms of space-time derivatives, which is guaranteed
by the projectivity. This restriction of the class of allowed field redefinition is anticipated to have an effect of
enlarging a class of nontrivial couplings to match Metsaev's classification of nontrivial vertices in HS theories in
any dimension. Indeed, there may be vertices that cannot be trivialised by a local field redefinition but can be compensated by a non-local one. For instance, this phenomenon was illustrated in \cite{PV,Prokushkin:1999xq} in the framework of $3d$ HS theory.

 As such, the proposed (S)HS theory in any
dimension has similarities with the $4d$ HS theory formulated in terms of spinor variables in
\cite{more}. In particular, only functions  linear in $\B$ may lead to local
vertices while all higher-order vertices are essentially nonlocal because of the star product in $\B * \B*\ldots$.
The terms linear in $\B$ on the other hand may lead to local interaction vertices analogously to the $4d$ model.
The derivation of the full list of vertices within the proposed model is
a challenging problem under investigation. Here we only speculate that the simplest
case of the theory of \cite{Vasiliev:2003ev} with no additional coupling constants
is somewhat analogous to the $4d$ HS self-dual  model originally proposed in \cite{more} as the $\bar\eta=0$ model. The similarity is
likely that in the both types of models there is no room for nontrivial current interactions. (Recall that for
instance stress tensor in the self-dual Yang-Mills theory vanishes.) Probably, for that reason the self-dual HS
theory (sometimes called chiral) admits specific fairly simple formulations   in four dimensions \cite{Sharapov:2022awp}-\cite{Sharapov:2022nps}, that admit a generalization to  any  dimension \cite{Didenko:2023vna,Didenko:2024zpd}.

An important problem for the future is to extend the obtained results
beyond the class of symmetric gauge fields.
There was a number of important contributions to this subject
in the literature. In particular, the light cone formulation of
the equations of motion of generic massless fields in $AdS_d$
\cite{meq} and actions for mixed symmetry  massless
fields in $AdS_5$ \cite{met} were constructed by Metsaev.
The unfolded formulation
of a $5d$ HS theory with mixed symmetry fields was studied in
\cite{SS2} in the sector of  Weyl 0-forms. The frame-like formulation
pioneered by Stanley Deser and the author was developed for particular
two-column mixed symmetry fields  in \cite{Alkalaev:2003hc}.
Extension of the flat  space results
to $(A)dS_d$ is not  straightforward for mixed symmetry fields because, generally,  irreducible massless systems in $AdS_d$ reduce to a collection
of irreducible massless fields in the flat limit \cite{meq,BMV}. Interesting results on
the incorporation of mixed symmetry fields were obtained in \cite{Alkalaev:2011zv}.
Though there are many other interesting papers on this deep subject
(see e.g. \cite{mixed,Burd})
we believe that the most promising generalization is related to so-called
Coxeter HS theories and their multiparticle extensions \cite{Vasiliev:2018zer,Tarusov:2025qfo}.

Once the conjecture of this paper is verified beyond the linearized approximation, it may have important implications for the paradigm of holographic
correspondence \cite{Maldacena:1997re,Gubser:1998bc,Witten:1998qj} via the example
of HS holography conjectured by Klebanov and Polyakov \cite{KP} and further developed in \cite{LP}-\cite{Giombi:2011kc}. Namely,
so far it is usually assumed that there are only two options for the HS
holography in higher dimensions, namely the $A$ and $B$-models dual to the free bosonic and fermionic boundary theories, or their supersymmetrization. (See,
however, \cite{Fei:2014yja}-\cite{Giombi:2019upv}.)
In that case there is no room for
the variety of coupling constants matching Metsaev's classification of the
independent vertices in the bulk.

Being obscure within the standard Klebanov-Polyakov HS holographic correspondence conjecture \cite{KP},
the origin of the broad class of HS couplings has better chances to be understood
within   an alternative HS holography conjecture of \cite{Vasiliev:2012vf}
suggesting that the duality is between
gravitational (HS) theories in $AdS_{d+1}$ and conformal theories in $d$ dimensions
interacting with conformal (HS) gravity on the boundary
(see also \cite{Diaz:2024kpr,Diaz:2024iuz}). If true it may resolve the
paradox in case the conformal HS gravity has as many  coupling
constants as the HS theory in  $AdS_{d+1}$. All this makes the detailed
analysis  of the vertices of the model proposed in this paper extremely
interesting. Note that going beyond the Klebanov-Polyakov conjecture might also
be interesting to reconsider the existing arguments for nonlocality of the HS theory that so far are all holography based on \cite{Bekaert:2015tva}-\cite{David:2020ptn}.

The last but not least is that the proposed BRST technique makes the formulation of the HS gauge theory closer to the BRST formulation of String Theory (see, {\it e.g., } \cite{Lust:1989tj} and references therein). As such, it
is anticipated to provide a promising tool for the unification of
HS theory and String Theory via association of the BRST operator $Q$ with $2d$ CFTs.

Though, naively, the BRST approach is of little use for the spinor formulation
of the $4d$ HS gauge theory because the factorisation of the
traceful components of the fields is automatically implemented within the spinor
formulation of \cite{more}, it is still useful for the analysis of Lorentz covariance of the theory \cite{GV}.

To summarize, we hope that this work sheds some new light on the structure of
HS gauge theory which is a fascinating subject exhibiting remarkable
properties, like, for instance, cancelation of quantum corrections even in the purely bosonic models   \cite{Giombi:2014yra,Beccaria:2015vaa}. HS theory is the field where Stanley Deser has made an outstanding contribution.

\section*{Acknowledgments}
I am grateful to  Vyatcheslav Didenko, Sergey Fedoruk,
Nikita Misuna, Anatoly Korybut and Alexandr Reshetnyak
for  useful comments, to Olga Gelfond, Mikhail Povarnin, Aleksandr Tarusov,  Yuri Tatarenko and Kirill Ushakov  for helpful discussions and many comments on the manuscript,
to Mirian Tsulaia for the correspondence,
to Evgeny Ivanov for a stimulating suggestion, that triggered
this work, and to Joseph Buchbinder for sharing his copy of the book
\cite{book}. I am grateful for hospitality to Ofer Aharony,
Theoretical High Energy Physics Group of Weizmann Institute of Science where the substantial
part of this work has been done. This work was supported by Theoretical Physics and
Mathematics Advancement Foundation “BASIS” Grant No 24-1-1-9-5.

\newcounter{appendix}
\setcounter{appendix}{1}
\renewcommand{\theequation}{\Alph{appendix}.\arabic{equation}}
\addtocounter{section}{1} \setcounter{equation}{0}
\renewcommand{\thesection}{\Alph{appendix}}
\addcontentsline{toc}{section}{\,\,\,\,\,\,\,Appendix A.   Compact spin-locality  }


\section*{Appendix A. Compact spin-locality}
\label{locc}

\subsection{Spin-local vertices in higher-spin theory}

Let us first  explain how the fibers  spin-locality
(called spinor spin-locality in the $4d$ HS models
\cite{Gelfond:2018vmi}) works.
HS vertices {can be put into the form}
\be\label{vert}
\Upsilon(\go,\go,\ldots, C,C,\ldots)=
F(Y, t_l^i ,\bt_l^i, p_l^{l},\bp_l^i )\go(Y_1)\ldots \go(Y_k) C(Y_{k+1})\ldots C(Y_n)\Big |_{Y_i=0}\,,
\ee
where
$$
t^i_l := V^A \frac{\p}{\p Y^A_{l i}}
$$
acts on the argument of the $l^{th}$ factor of $\go$ and
\be
p^i_l:= V^A \frac{\p}{\p Y^A_{l i}}
\ee
acts on the argument of the $l^{th}$ factor of $C$.

$$
\bp^i_{la} := \f{\p}{\p Y^a_{l i}}
$$
and
$$
\bt^i_{la} := \f{\p}{\p Y^a_{l i}}
$$
act on the Lorentz parts $Y^a_i$ of the arguments $Y^A_i$ of the $l^{th}$ factors of $C$ and $\go$, respectively.
The function $F(Y, t_l^i ,\bt_l^i, p_i^{l},\bp_l^i )$
depends on various Lorentz invariant and $sp(2)$ invariant contractions of $t,\bt,p$ and $\bp$.
The dependence on $t$ and $\bt$ does not affect spin locality since, for any given spin, the
one-form $\go$ contains a finite number of derivatives of the dynamical HS field (see e.g. \cite{Bekaert:2004qos}).
Also, one can see that the terms containing $\bp^i_{la}\bp^j_{n b} \gvep_{ij}$  do not affect
spin locality since the number of the Lorentz indices in the second row of the two-row Young tableaux
equals to spin and hence is bounded. Moreover the respective vertices are spin-local compact \cite{Vasiliev:2022med} because the
increase of the spin $s_l$  of some field $C$ would imply the increase of the number of indices in the first
row associated with $C(Y_l)$. Since the number of uncontracted indices (\ie the degree in $Y$)
is limited by the spin of the vertex $s_0$, $s_l$ cannot be larger than the number of indices in all second
rows of the Young tableaux associated with
the fields $C$, \ie the sum of spins in the vertex. As a result, whether the number of derivatives in the
vertex is finite or not is controlled by the dependence of $F(Y, t_l^i ,\bt_l^i, p_i^{l},\bp_l^i )$ on
$p_i^l p^{i n}$ for various $l\neq n$ (note that $p_i^l p^{i n}=-p_i^n p^{i l}$).  If the dependence of $F$ on
$p_i^l p^{i n}$ is polynomial for all pairs of $l,n$ (\ie pairs of the factors of $C$), the vertex is spin-local and, in fact, spin-local-compact by virtue of the
arguments analogous to those presented above. Otherwise, the vertex is non-local.

Following \cite{Vasiliev:2022med},  we recall   peculiarities of  the notions of locality and non-locality in field theories like HS gauge theory, that contain higher derivative
interaction vertices for infinite towers of fields
of different spins \cite{Bengtsson:1983pd, Berends:1984rq, FV1}.

 Since the order of maximal  derivatives
  in a HS vertex $V(s_1,s_2,s_3)$
 for three fields  with spins $s_1,s_2,s_3$  increases with
 involved spins \cite{Metsaev:2007rn}, the number of derivatives in the
  theory  is unbounded
 once all spins are involved. Such a theory is non-local in the standard sense.
 However, there are more options to  distinguish between.

\subsection{Interactions}

 Let some system describe  fields $\phi_s^A$ characterized by
 quantum numbers called spin $s$ and some Lorentz indices $A$ like
 tensor, spinor, etc. Consider field equations of the  form
$$
E_{A_0,s_0}(\p ,\phi) =\sum_{k=0,l=1}^\infty
 a^{n_1\ldots n_k}_{A_0\,A_1\ldots A_l} (s_0,s_1,s_2,\ldots\,, s_l)
 \p_{n_1}\ldots \p_{n_k}\phi_{s_1}^{A_1}\ldots\phi_{s_l}^{A_l}=0\,.
$$
Here derivatives $\p_n:= \frac{\p}{\p x^n}$ may hit any of the fields $\phi_{s_k}^{A_k}$ with
$s_0$ being the spin of the field on which the linearized  equation is imposed.
{Locality of the equations can be treated  perturbatively, \ie
independently at every order $l$.
In {usual perturbatively local} field theory the total number of
derivatives is limited at any order $l$ by some $k_{max}(l)$:}
\be
\label{loc}
a^{n_1\ldots n_k}_{A_0\ldots A_l} (s_0,s_1,s_2,\ldots s_l) =0 \quad \mbox{at}\quad
k>k_{max}(l)\,.
\ee

This condition can be relaxed to {\it space-time spin-locality} condition
\be
\label{sloc}
a^{n_1\ldots n_k}_{A_0\ldots A_l} (s_0,s_1,s_2,\ldots s_l) =0 \quad \mbox{at}\quad
k>k_{max}(s_0,s_1,s_2,\ldots s_l)
\ee
with some $k_{max}(s_0,s_1,s_2,\ldots s_l)$ depending on the spins in the vertex.
In the theories with the finite number of fields where $s$ can take at most a
finite number of values, the conditions (\ref{loc}) and (\ref{sloc}) are
equivalent. However in the HS-like models, with infinite towers of spins,
the locality and spin-locality restrictions differ. Both  types of theories
have to be distinguished from the genuinely non-local ones in which
there exists such a subset of spins $s_0,s_1,s_2,\ldots s_l$ that (\ref{sloc})
is not true, \ie no finite $k_{max}(s_0,s_1,s_2,\ldots s_l)$  exists at all.

The relaxation of the class of local field theories with the finite number of
fields to the spin-local class is the simplest appropriate  for the
models involving infinite towers of fields. However, it makes sense to further
specify the concept of spin-local vertices.

Following \cite{Vasiliev:2022med}, we call a spin-local vertex {\it compact} if
$ a^{n_1\ldots n_k}_{A_0\,A_1\ldots A_l} (s_0,s_1,s_2,\ldots\,,s_k +t_k\,,\ldots, s_l)=0$ at
$t_k> t_k^0$ with some $t_k^0$
 for any $0\leq k\leq l$ and {\it non-compact}
otherwise. (Note that this is compactness in the space of spins - not space-time.)
 In HS theory both types of vertices are present. (For more detail see
 \cite{Vasiliev:2022med} and references therein.)

\subsection{Field redefinitions}
{A class of perturbatively local theories with finite sets of fields is invariant under
 perturbatively local field redefinitions}
 \be
 \label{fr}
\phi^B_{s_0}\to \phi^B_{s_0} +\delta \phi^{B}_{s_0}\q \delta \phi^{B}_{s_0} = \sum_{k=0,l=1}^\infty
b^B{}^{n_1\ldots n_k}_{A_1\ldots A_l}(s_0,s_1,\ldots, s_l)
 \p_{n_1}\ldots \p_{n_k}\phi_{s_1}^{A_1}\ldots\phi_{s_l}^{A_l}
\ee
{with at most finite number of non-zero coefficients $b^B{}^{n_1\ldots n_k}_{A_1\ldots A_l}(s_0,s_1,\ldots, s_l)$
 at any given order.}
Note that application of a non-local perturbative
field redefinition to a local field theory makes it seemingly non-local.

Once the (spin-)local frame of a model is known, the next question
 is what is the proper class of field redefinitions that leave
the form of vertices perturbatively local or spin-local? In field theories
with a finite number of fields the answer is that these are
perturbatively local field redefinitions involving a finite number of
derivatives at every order.

In the models with infinite sets of fields the situation is more
subtle. Naively one might think that appropriate field redefinitions in spin-local theories
 are also spin-local. This is not necessarily  true,
however, because of the infinite summation over the spin $s_p$ of the redefined field in the modified vertex\,,
\bee
\delta  E_{A_0,s_0}(\p ,\phi) \ls&&=\sum_{{s_p}=0}^\infty
\sum_{p,k,k'=0,l,l'=1}^\infty\ls
 a^{n_1\ldots n_k}_{A_0\,A_1\ldots A_l} (s_0,s_1,s_2,\ldots,s_p,\ldots\,, s_l)\times
\\
&&  \ls\ls\ls\ls\times \p_{n_1}\ldots \p_{n_k}\phi_{s_1}^{A_1}\ldots\phi^{A_{p-1}}_{s_{p-1}}
 \phi^{A_{p+1}}_{s_{p+1}}
\ldots  \phi_{s_l}^{A_l}\times\\&& \ls\times
b^{A_p}{}^{m_1\ldots m_{k'}}_{B_1\ldots B_{l'}}(s_p,s_{l+1},\ldots, s_{l+l'})
 \p_{m_1}\ldots \p_{m_{k'}}\phi_{s_{l+1}}^{B_1}\ldots\phi_{s_{l+l'}}^{B_{l'}}
\,.\nn
\eee
If the vertex and field redefinition were spin-local the result of such a
field redefinition can still be non-local and even ill-defined because an
infinite number of terms with the same field pattern and any number of
derivatives may result from the  terms with different
$s_p$.

This problem is avoided provided that  the field redefinition (\ref{fr})
is spin-local-compact in which case the summation over $s_p$ is always
finite and the modified vertex is both well-defined and spin-local. Thus, in the
spin-local theories with  infinite sets of fields a proper class
is represented by spin-local-compact field redefinitions.  An output
 of this analysis is that the gauge transformations
including the ideal factorization ones  have to be spin-local-compact.

\setcounter{appendix}{2}
\renewcommand{\theequation}{\Alph{appendix}.\arabic{equation}}
\addtocounter{section}{1} \setcounter{equation}{0}
\renewcommand{\thesection}{\Alph{appendix}}
\addcontentsline{toc}{section}{\,\,\,\,\,\,\,Appendix B.  $sp(2)$ factorisation versus invariance }

\section*{Appendix B. $sp(2)$ factorisation versus invariance}
In this section we sketch the analysis of the potential conflict
of $sp(2)$ factorisation versus invariance in the BRST free setup.

The  fermionic fields $a_{12}$ and $a_{21}$ have to respect both
the $t_i$ invariance and the $t_i$ factorization conditions.
Because $t_i$ contains a factor of the $\Pi_2$ (\ref{p12}) while $a_{12}$ and
$ a_{21}$ are proportional to $\Theta$ and $\bar\Theta$, respectively, the former implies (\ref{tistar})
\be
\label{tistar=}
t_i * a_{12} (Y,\phi_+)*\delta^m (\phi_-)=0\,,\q \delta^m (\phi_-)*a_{21} (Y,\phi_-) *t_i =0\,.
\ee
One can see that these conditions imply $\gamma$-transversality of the respective
spinor-tensors, that makes them Lorentz irreducible. However, in addition one has to factor out the
terms of the form (\ref{tijk})
\be
\label{tijk=}
t_{\Lambda \Phi}*b^{\Lambda \Phi}_{12} \sim 0\,
\ee
provided that $t_{\Lambda \Phi}*b^{\Lambda \Phi}_{12}$ is $osp(1,2)$ invariant. Here is a potential subtlety: while the part of these conditions associated with
$t_{ij}$  implies the tracelessness of the spinor tensor in the $Y^A_i$
variables, the second one again contains the
$\gamma$-traces. That is fine at the linearized level but may be
problematic beyond if the latter condition  and (\ref{tistar=})  are deformed
differently by the nonlinear corrections. Here we explain the
mechanism guaranteeing that this does not happen, \ie that the both conditions are equivalent.

Indeed, let $\tau_i$ and $\tau_{ij}$ denote the $osp(1,2)$ generators
in the nonlinear theory as defined in Section \ref{tnon}. Let
\be
\ta_i* \varphi_{12}^i
\ee
be in the ideal to be factored out. This demands that it must be $\ta_j$ invariant, \ie
\be
\label{ttv}
\ta_j *\ta_i *\varphi_{12}^i =0\,.
\ee
On the other hand,  relation (\ref{ttL}) applied to the nonlinear generators $\tau$ implies
\be
L * \tau_i* \varphi_{12}^i = \tau_{ij} *\tau^j*\varphi^i_{12}\,
\ee
while with the help of (\ref{L}), (\ref{ttv})  yields
\be
L * \ta_i* \varphi_{12}^i =2 \ta_i* \varphi_{12}^i\,
\ee
and, hence,
\be
\label{tij}
 \ta_i* \varphi_{12}^i = \half \ta_{ij} *\ta{}^j*\varphi^i_{12}\,.
\ee
Therefore, for $t_i$ invariant elements in the sector $12$, the $\tau_i$ factorisation  is a consequence of the $\tau_{ij}$ factorisation. Analysis of the sector $21$ is analogous.

Here one has to be careful
however when analysing whether or not the terms (\ref{tij}) are indeed factorisable as belonging to the appropriate
functional class. If not, this implies that the factorisation by this mechanism does not occur,
that may imply that some additional vertices involving $\tau_i$ may survive.
A related point is that within the naive factorization, the
$t_{ij} *t^{ij}$ term on the \rhs of (\ref{LL}) drops out. It can however contribute to nontrivial vertices if  some of such terms are beyond the appropriate projectively-compact spin-local class (see Appendix A).

In any case, we believe that the BRST approach developed in the paper resolves this
problem in a much nicer way being preferable for the analysis.

\end{document}